\documentclass[journal]{IEEEtran}

\usepackage{amsthm}
\usepackage{float}
\usepackage{cite}										
\usepackage{amsmath}									
\usepackage{graphicx}
\usepackage{epstopdf}
\usepackage{float}
\usepackage{bm,upgreek}
\usepackage{amsfonts}
\usepackage{enumitem}
\usepackage{mathtools}
\usepackage{multirow}
\usepackage{booktabs}
\usepackage[subnum]{cases}
\usepackage{setspace}
\usepackage{color}
\usepackage{blkarray, bigstrut} 
\usepackage[caption=false,labelformat=simple]{subfig}

\usepackage{xcolor}
\usepackage{pgfplots}
\pgfplotsset{compat=1.5}
\usepackage{pgfplotstable}
\usepgfplotslibrary{statistics}
\pgfplotsset{grid style={dotted,gray}}
\usepackage{tikz}
\usetikzlibrary{math}
\usepackage{hyperref}

\usepackage{mathtools,nccmath}
\usepackage[ruled]{algorithm}
\usepackage{algpseudocode,algorithmicx}

\algnewcommand{\LineComment}[1]{\State \(\#\) #1}

\allowdisplaybreaks

\ifCLASSINFOpdf
\else
\fi
\pgfplotsset{legend image with text/.style={legend image code/.code={%
\node[anchor=west, align=right] at (0.0cm,0cm) {#1};}},}

\makeatletter
\def\myline{\pgfutil@ifnextchar[{\my@line}{\my@line[]}}%
\def\my@line[#1](#2)(#3){%
\tikz[overlay] \draw[#1]  (#2)--(#3); 
}%
\algrenewcommand\algorithmicindent{1.0em}%

\makeatletter
\renewcommand{\ALG@beginalgorithmic}{\small}
\makeatother



\usepackage{etoolbox}
\BeforeBeginEnvironment{algorithmic}{\kern 0.3\baselineskip}
\AfterEndEnvironment{algorithmic}{\kern -0.1\baselineskip}

\newtheorem{theorem}{Theorem}

\makeatletter
\renewcommand{\ALG@beginalgorithmic}{\footnotesize}
\makeatother

\pgfplotsset{
    box plot/.style={
        /pgfplots/.cd,
        black,
        only marks,
        mark=-,
        mark size=\pgfkeysvalueof{/pgfplots/box plot width},
        /pgfplots/error bars/y dir=plus,
        /pgfplots/error bars/y explicit,
        /pgfplots/table/x index=\pgfkeysvalueof{/pgfplots/box plot x index},
    },
    box plot box/.style={
        /pgfplots/error bars/draw error bar/.code 2 args={%
            \draw  ##1 -- ++(\pgfkeysvalueof{/pgfplots/box plot width},0pt) |- ##2 -- ++(-\pgfkeysvalueof{/pgfplots/box plot width},0pt) |- ##1 -- cycle;
        },
        /pgfplots/table/.cd,
        y index=\pgfkeysvalueof{/pgfplots/box plot box top index},
        y error expr={
            \thisrowno{\pgfkeysvalueof{/pgfplots/box plot box bottom index}}
            - \thisrowno{\pgfkeysvalueof{/pgfplots/box plot box top index}}
        },
        /pgfplots/box plot
    },
    box plot top whisker/.style={
        /pgfplots/error bars/draw error bar/.code 2 args={%
            \pgfkeysgetvalue{/pgfplots/error bars/error mark}%
            {\pgfplotserrorbarsmark}%
            \pgfkeysgetvalue{/pgfplots/error bars/error mark options}%
            {\pgfplotserrorbarsmarkopts}%
            \path ##1 -- ##2;
        },
        /pgfplots/table/.cd,
        y index=\pgfkeysvalueof{/pgfplots/box plot whisker top index},
        y error expr={
            \thisrowno{\pgfkeysvalueof{/pgfplots/box plot box top index}}
            - \thisrowno{\pgfkeysvalueof{/pgfplots/box plot whisker top index}}
        },
        /pgfplots/box plot
    },
    box plot bottom whisker/.style={
        /pgfplots/error bars/draw error bar/.code 2 args={%
            \pgfkeysgetvalue{/pgfplots/error bars/error mark}%
            {\pgfplotserrorbarsmark}%
            \pgfkeysgetvalue{/pgfplots/error bars/error mark options}%
            {\pgfplotserrorbarsmarkopts}%
            \path ##1 -- ##2;
        },
        /pgfplots/table/.cd,
        y index=\pgfkeysvalueof{/pgfplots/box plot whisker bottom index},
        y error expr={
            \thisrowno{\pgfkeysvalueof{/pgfplots/box plot box bottom index}}
            - \thisrowno{\pgfkeysvalueof{/pgfplots/box plot whisker bottom index}}
        },
        /pgfplots/box plot
    },
    box plot median/.style={
        /pgfplots/box plot,
        /pgfplots/table/y index=\pgfkeysvalueof{/pgfplots/box plot median index},
        thin,black
    },
    box plot width/.initial=1em,
    box plot x index/.initial=0,
    box plot median index/.initial=1,
    box plot box top index/.initial=2,
    box plot box bottom index/.initial=3,
    box plot whisker top index/.initial=4,
    box plot whisker bottom index/.initial=5,
}

\newcommand{\boxplot}[2][]{
    \addplot [box plot median,#1] table {#2};
    \addplot [forget plot, box plot box,#1] table {#2};
    \addplot [forget plot, box plot top whisker,#1] table {#2};
    \addplot [forget plot, box plot bottom whisker,#1] table {#2};
}

\pgfplotsset{my legend/.style={
    legend image code/.code={
        \fill [#1] (0cm,-0.1cm) rectangle (0.3cm,0.1cm);
    },
}}

\algnewcommand\algorithmicforeach{\textbf{for each}}
\algdef{S}[FOR]{ForEach}[1]{\algorithmicforeach\ #1\ \algorithmicdo}

\include{acronyms}

\definecolor{coal}{HTML}{56CBF4}
\definecolor{oil}{HTML}{4876BB}
\definecolor{gas}{HTML}{86CA88}
\definecolor{nuclear}{HTML}{15AEA1}
\definecolor{hydro}{HTML}{F9F162}
\definecolor{geothermal}{HTML}{4F4D2E}
\definecolor{wind}{HTML}{F3744E}
\definecolor{solar}{HTML}{A186BE}
\definecolor{biofuels}{HTML}{AB6AAC}

\begin{document}

\newlist{myitemize}{itemize}{3}
\setlist[myitemize,1]{label=\textbullet,leftmargin=6.5mm}

\title{Distributed Inference over Linear Models using Alternating Gaussian Belief Propagation}

\author{Mirsad Cosovic,~\IEEEmembership{Member,~IEEE,}
        Dragisa Miskovic,~\IEEEmembership{Member,~IEEE,}
        Muhamed Delalic,~\IEEEmembership{Member,~IEEE,}
        Darijo Raca,~\IEEEmembership{Member,~IEEE,}
        Dejan Vukobratovic,~\IEEEmembership{Senior Member,~IEEE}

\thanks{M. Cosovic is with Faculty of Electrical Engineering, University of Sarajevo, Bosnia and Herzegovina, and the Institute for Artificial Intelligence Research and Development of Serbia (e-mail: mcosovic@etf.unsa.ba); D. Miskovic is with the Institute for Artificial Intelligence Research and Development of Serbia (e-mail: dragisa.miskovic@ivi.ac.rs); M. Delalic and D. Raca are with Faculty of Electrical Engineering, University of Sarajevo, Bosnia and Herzegovina (e-mail: muha.delalic@gmail.com, draca@etf.unsa.ba); D. Vukobratovic is with Faculty of Technical Sciences, University of Novi Sad, Serbia, (email: dejanv@uns.ac.rs).}
\thanks{This paper has received funding from the European Union's Horizon 2020 research and innovation programme under Grant Agreement number 856967.}}

\markboth{}%
{Shell \MakeLowercase{\textit{et al.}}: Bare Demo of IEEEtran.cls for IEEE Journals}

\maketitle

\begin{abstract}
We consider the problem of maximum likelihood estimation in linear models represented by factor graphs and solved via the Gaussian belief propagation algorithm. Motivated by massive internet of things (IoT) networks and edge computing, we set the above problem in a clustered scenario, where the factor graph is divided into clusters and assigned for processing in a distributed fashion across a number of edge computing nodes. For these scenarios, we show that an alternating Gaussian belief propagation (AGBP) algorithm that alternates between inter- and intra-cluster iterations, demonstrates superior performance in terms of convergence properties compared to the existing solutions in the literature. We present a comprehensive framework and introduce appropriate metrics to analyse AGBP algorithm across a wide range of linear models characterised by symmetric and non-symmetric, square, and rectangular matrices. We extend the analysis to the case of dynamic linear models by introducing dynamic arrival of new data over time. Using a combination of analytical and extensive numerical results, we show the efficiency and scalability of AGBP algorithm, making it a suitable solution for large-scale inference in massive IoT networks.
\end{abstract}


\IEEEpeerreviewmaketitle

\section{Introduction}
Recent years have witnessed breakthroughs in embedded hardware technologies, paving the way for the emergence of powerful, low-cost and low-power, multi-functional sensor devices. These devices have capability of acquiring and processing information obtained from the environment in its close proximity. The sensed information about the environment can be propagated from almost anywhere via, e.g., massive machine-type communication service of the fifth-generation (5G) mobile cellular networks to the edge or cloud computing nodes for data processing, thus establishing \emph{massive internet of things} (IoT) networks~\cite{guo2021enabling}. The ubiquitous massive IoT networks support the creation of a large-scale, cost-effective, and spatially interconnected smart environments empowered by big data analysis~\cite{lv2021big}. The rapid development of 5G and beyond-5G networks builds the foundation of a unified infrastructure for large-scale information acquisition, communication, storage, and computing~\cite{cosovic20175g}. 

The distinctive feature of massive IoT networks is the availability of vast information distributed across geographically distant areas, posing limitations for its timely and efficient sharing and joint processing. 
For this reason, processing this information in a centralised manner by transmitting and processing all the data at one place, represents an impractical approach in practice~\cite{shikhar}. Storing and processing data at one place has a serious implication on data privacy and security, issues traditionally overlooked in a centralised approach. One approach to lessen these concerns is to process data in decentralised manner, where each sensor preprocess data before sending the modified data to the central entity for final decision achieving good real-time performances~\cite{9258964, 6971234}. With proliferation of edge computing capacities, a distributed processing approach is emerging as a promising and viable alternative. Distributed edge computing can overcome the issues associated with the centralised approach, conforming with the distributed nature of massive IoT networks~\cite{liu2020toward}. 

Processing information in massive IoT networks often represents an instance of statistical estimation problems in large-scale systems of random variables. The maximum likelihood estimation principle underpins majority of practical estimators~\cite[Sec.~7.1]{kay}. In the literature, numerous examples exist including channel~\cite{du} and position estimation~\cite{wymeersch} in communication networks, state estimation in electric power systems~\cite{cosovic20175g}, and device tracking using mobile sensors~\cite{olfati}. Matrix decomposition techniques such as lower–upper or orthogonal factorisation \cite{korres} have been recognised as the foundations for solving linear models using maximum likelihood estimation in both centralized and distributed manner~\cite{wang}. For these techniques, the time complexity increases exponentially with the number of state variables, making it impractical for large-scale inference. 

To overcome this drawback, iterative algorithms provide a feasible alternative for large-scale inference~\cite{marelli}. Alternatively to well-known algorithms such as Jacobi and Gauss-Seidel, one can apply an inference algorithms in graphical models, such as the \emph{Gaussian belief propagation} (GBP) algorithm~\cite{loeliger}, known to converge more rapidly for large-scale systems \cite{bickson, ortiz2021visual}. Another advantage of using the GBP algorithm is its easy \emph{distribution} across disjoint computation nodes, where disjoint parts of a graphical model are processed by a set of interconnected computation nodes (e.g., edge computing nodes in 5G network), while maintaining the complexity that scales \emph{linearly} per iteration~\cite{Azhang, bin}. 

Most existing solutions that leverage the GBP algorithm~\cite{hu, sui, tai, cosovictra} in a distributed setting are based on synchronous scheduling of GBP iterations among computing nodes (clusters), thus relying on synchronous message exchange between disjoint clusters. In this paper, we challenge the efficiency of such an approach, and propose a novel alternating GBP (AGBP) algorithm for the distributed settings. In particular, AGBP employs alternating scheduling, where number of intra-cluster iterations (i.e., synchronous message exchanges within each of the clusters) are done interchangeably with a number of inter-cluster iterations (synchronous message exchanges between the clusters). Using analytical results and extensive numerical examples, we argue that alternating scheduling is an efficient approach for solving large-scale inference problems in massive IoT networks under communication constraints.

The impact of our work reflects in two key outcomes: improved convergence rate and improved probability of convergence for the AGBP as compared to the traditional synchronous GBP. The contributions of this work are summarised as follows:
\begin{itemize}
    \item We show a significant improvement in the inference time of the AGBP algorithm compared to the synchronous GBP.
    \item We show through experiments that the convergence rate is independent of number of clusters and is mainly driven by the \emph{sparsity} of the clusters a factor graph is segmented into. 
    \item Applying AGBP in a dynamic scenario, where new observations and state estimates are evolved continuously over time, causes additional reduction of iterations until convergence.
    \item In scenarios where both alternating and synchronous GBP algorithms fail to converge, applying damping technique (see Appendix D for details) is more beneficial for the AGBP algorithm.
\end{itemize}

In summary, the above remarks make the AGBP algorithm a scalable and efficient method for distributed inference in massive IoT networks, in particular for the cases with dynamic arrival of new observations over time.

The rest of the paper is organised as follows. Section II establishes the theoretical basis and presents the GBP background. In Section III, we apply the GBP algorithm to the distributed architecture, with a focus on reducing communication delay by introducing an alternating message scheduling, which forms the core of the AGBP algorithm. Section IV presents the numerical results of various testing scenarios, and Section V outlines the key findings and conclusions of the paper.

\section{Probabilistic Inference in Linear Models}
In this section, we review the GBP algorithm to solve a generic linear model \cite{loeliger}. We set the solution of a probabilistic inference problem using GBP in the context of solving a maximum likelihood estimation problem. Further, to efficiently solve these models under non-stationary conditions, we expand our setup to a dynamic GBP algorithm. 

\subsection{Maximum Likelihood Estimation}
We consider a linear model described by a noisy system of linear equations:
\begin{equation}
    \mathbf{z} = \mathbf{h}(\mathbf{x}) + \mathbf{u},
    \label{eqn:linear_model}
\end{equation}
where $\mathbf{x} = [x_1, \dots, x_n]^T$ represents a vector of state variables, $\mathbf{z} = [z_1,\dots, z_m]^T$ is an observation vector, while $\mathbf{u} = [u_1, \dots, u_m]^T$ is a vector of uncorrelated observation errors, where $u_i \sim \mathcal{N}(0, v_{i})$ follows a zero-mean \emph{Gaussian distribution} with variance $v_{i}$. The vector $\mathbf{h}(\mathbf{x}) = [h_1(\mathcal{X}_1), \dots, h_m(\mathcal{X}_m)]^T$ comprises a collection of $m$ linear equations, where $\mathcal{X}_i \subseteq \mathcal{X}$, $i=1,\dots,m$, represents a subset of the set of state variables $\mathcal{X} = \{x_1, \dots, x_n \}$. Under these conditions, the probability density function associated with the $i$-th observation can be written as follows: 
\begin{equation}
    \mathcal{N}(z_i | \mathcal{X}_i, v_i) = \frac{1}{\sqrt{2 \pi v_i}}\mathrm{exp} \Bigg\{- \frac{[z_i - h_i(\mathcal{X}_i)]^2}{2v_i} \Bigg\}.
    \label{eqn:gauss_probability}
\end{equation}

The maximum likelihood solution or \emph{estimate} can be obtained by solving the following optimisation problem:
\begin{equation}
    \hat{\mathbf{x}} = \underset{\mathbf{x}}{\mathrm{arg max}}\; \mathcal{N}(\mathbf{z} | \mathbf{x}, \mathbf{\Sigma}) = \underset{\mathbf{x}}{\mathrm{arg max}} \displaystyle\prod_{i=1}^{m} \mathcal{N}(z_i | \mathcal{X}_i, v_i),
    \label{eqn:optimisation_problem}
\end{equation}
where the positive definite covariance matrix $\bm{\Sigma} \in \mathbb{R}^{m \times m}$ contains the corresponding variances as diagonal entries.  Note that each entry of estimate $\hat{\mathbf{x}}$ represents a mean of the marginal likelihood (distribution) of the corresponding variable from the set $\mathcal{X}$, thus it can be also obtained via efficient marginalisation using GBP algorithm (see Section \ref{sec2b}).

In many technical fields such as statistics, signal processing, and control theory, linear models \eqref{eqn:linear_model} are typically overdetermined, with $m \geq n$, written in matrix form:
\begin{equation}
    \mathbf{z} = \mathbf{H} \mathbf{x} + \mathbf{u},
    \label{eqn:linear_model_matrix}
\end{equation}
where the coefficient matrix $\mathbf{H} \in \mathbb{R}^{m \times n}$ has a full column rank. Then, starting from \eqref{eqn:optimisation_problem}, the maximum likelihood solution can be obtained by solving the weighted least-squares equation \cite[Sec.~7.8]{kay}:
\begin{equation}
    \hat{\mathbf{x}} = (\mathbf{H}^T \mathbf{\Sigma}^{-1} \mathbf{H})^{-1} \mathbf{H}^T \mathbf{\Sigma}^{-1} \mathbf{z}.
    \label{eqn:wls}
\end{equation}

We focus on a \emph{fixed} and \emph{static} linear models \eqref{eqn:linear_model_matrix}, where the $i$-th sensor provides an observation $z_i$ modeled as $\mathcal{N}(z_i | \mathcal{X}_i, v_i)$, defined in \eqref{eqn:gauss_probability}. In addition to static scenarios, we also analyse \emph{dynamic} linear models in which the observation value $z_i$ and the corresponding variance $v_i$ change their values in discrete time instances. The described dynamic scenario can be solved at each time instance using either \eqref{eqn:wls}, or applying the sliding-window recursive least-squares algorithm with forgetting factors \cite{zhang}. The former approach represents an inefficient method because the system has to be solved from scratch at each time instance, while the latter approach efficiently solves dynamical systems, recursively updating the estimate $\hat{\mathbf{x}}$ according to new data.

\subsection{Gaussian Belief Propagation}\label{sec2b}

The locality of sensor observations implies sensing data in its close proximity, allowing likelihood function $\mathcal{N}(\mathbf{z} | \mathbf{x}, \mathbf{\Sigma})$ to be factorised in such a way that every function $\mathcal{N}(z_i | \mathcal{X}_i, v_i)$ contains small subsets of state variables $\mathcal{X}_i$. The resulting linear model has the property of \emph{sparsity}, where each sensor observes a function $h_i(\mathcal{X}_i)$ of a small number of state variables $\mathcal{X}_i$. This fact motivates solving the maximum likelihood estimation \eqref{eqn:optimisation_problem} for both static and dynamic scenarios in a \emph{distributed framework} in a scalable and efficient way using the GBP algorithm. The GBP algorithm efficiently computes the marginal distributions of the state variables from the set $\mathcal{X}$. We use graphical models called \emph{factor graphs} to represent the linear model \eqref{eqn:linear_model_matrix}, while we use the GBP algorithm to obtain a distributed solution of the optimisation problem \eqref{eqn:optimisation_problem}. The factor graph is a bipartite graph that describes the structure of factorisation \eqref{eqn:optimisation_problem} using a graph-based representation of probability density functions using \emph{variable} and \emph{factor nodes} connected by edges \cite[Ch.~8]{bishop}. This case is widely applicable to many real-world scenarios, benefiting from the fact that all mathematical operations during GBP iterations result in Gaussian distributions. 

Applying the GBP algorithm to linear models \eqref{eqn:linear_model_matrix} requires forming a factor graph with a structure consisting of the set of factor nodes $\mathcal{F}=\{f_1,\dots,f_m\}$, where each factor node $f_i$ represents a local function $\mathcal{N}(z_i | \mathcal{X}_i, v_i)$, connecting the set of variable nodes $\mathcal{X}$. The factor node $f_i$ connects to the variable node $x_j$ if and only if $x_j \in \mathcal{X}_i$ \cite{kschischang}. The set of factor nodes $\mathcal{F}$ can be divided according to the degree of the factor node $f_i$ defined by $\text{deg}(f_i) = |\mathcal{X}_i|$. We define the set of \emph{branch} factor nodes $\mathcal{B}$, where $f_i \in \mathcal{B}$ if $\text{deg}(f_i) > 1$, and the set of \emph{leaf} factor nodes $\mathcal{L}$, where $f_i \in \mathcal{L}$ if $\text{deg}(f_i) = 1$. Observing the matrix $\mathbf{H}$ leads to another interpretation of the factor graph, where each row of the matrix $\mathbf{H}$ corresponds to one factor node, while the columns define the variable nodes. A factor node connects to a variable node if and only if the corresponding coefficient of the matrix row is nonzero. Furthermore, if the $i$-th row contains only one nonzero element, the corresponding row defines the leaf factor node $f_i \in \mathcal{L}$, otherwise the row defines the branch factor node $f_i \in \mathcal{B}$.

In general, the GBP algorithm passes two types of messages along the edges of the factor graph: 
\begin{itemize}
    \item a variable node $x_j \in \mathcal{X}$ to a factor node $f_i \in \mathcal{B}$ message $\mu_{x_j \to f_i}(x_j)$, and
    \item a factor node $f_i \in \mathcal{F}$ to a variable node $x_j \in \mathcal{X} $ message $\mu_{f_i \to x_j}(x_j)$.
\end{itemize}
In GBP algorithm, both variable and factor nodes in a factor graph process incoming messages and calculate outgoing messages, where an output message on any edge depends on incoming messages from all other edges connected to a particular node. Details of implementation of GBP algorithm including message construction and computation can be found in Appendix A. For most of practical applications, factor graph representation includes cycles, requiring use of a \emph{loopy} GBP. Loopy GBP is an iterative algorithm, with the iteration index $k=\{0,1,\dots\}$, requiring a message-passing scheduling. The commonly used message-passing scheduler is the \emph{synchronous scheduling} \cite{elidan}, where messages from variable to factor nodes $\bm{\mu}_{x} = [\mu_{x_j \to f_i}(x_j)]$, $f_i \in \mathcal{B}$, $x_j \in \mathcal{X}$, and from factor to variable nodes $\bm{\mu}_{f} = [\mu_{f_i \to x_j}(x_j)]$, $f_i \in \mathcal{B}$, $x_j \in \mathcal{X}$, are updated in parallel in respective half-iterations\footnote{If necessary, to denote a matrix $\mathbf{A}$ or a vector $\mathbf{a}$, we write $\mathbf{A} = [a_{ij}]$ or $\mathbf{a} = [a_i]$, where $a_{ij}$ and $a_i$ represent generic elements.}. Note that the messages from leaf factor nodes $\mu_{f_i \to x_j}(x_j)$, $f_i \in \mathcal{L}$, $x_j \in \mathcal{X}$ remain constant in all iterations $k$. 

The synchronous GBP, running in the \emph{centralised} framework, starts with the initial values of factor node to variable node messages $\bm{\mu}^{(0)}_{f}$. The first iteration $k=1$ proceeds with the calculation of the messages from the variable nodes to the factor nodes $\bm{\mu}^{(1)}_{x}$, followed by a calculation of the messages from the factor nodes to the variable nodes $\bm{\mu}^{(1)}_{f}$. At the end of the first iteration, the GBP produces its first estimate $\hat{\mathbf{x}}^{(1)}$. The time required for the calculation of the estimates $\hat{\mathbf{x}}$ consists of variable initialisation duration, duration of variable-node processing $\tau_x$, and the duration of factor-node processing $\tau_f$. Assuming that the initialisation duration is negligible, GBP produces an estimate $\hat{\mathbf{x}}^{(\nu)}$ that satisfies a given convergence condition (e.g., that $\hat{\mathbf{x}}^{(\nu)}$ is sufficiently close the exact solution $\hat{\mathbf{x}}$ under a given estimation accuracy metric) after $k=\nu$ iterations at the time instant $\nu\tau_{m}$, $\tau_{m} = \tau_x + \tau_f$. It is important to note that the computation time $\tau_x$ results from evaluating \eqref{BP_vf_mean_var} and \eqref{BP_marginal_mean_var}, whereas the time $\tau_f$ is determined by \eqref{BP_fv_mean_var}.

\section{Distributed Probabilistic Inference in the Linear Models}  
\begin{figure}[ht]
	\centering
	\includegraphics[width=9.2cm]{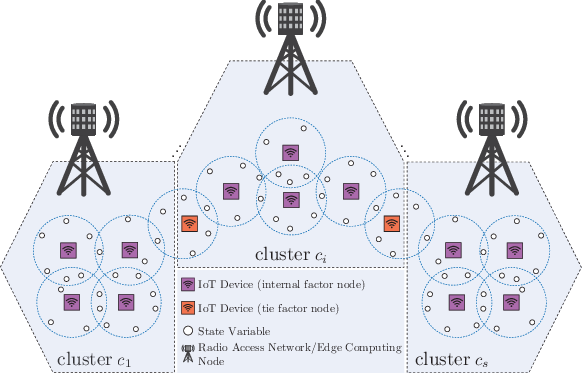}
	\caption{The architecture of a distributed IoT network supported by both radio access network and edge computing nodes.}
	\label{fig:bird}
\end{figure}

For massive IoT networks, the resulting system could be segmented and allocated to different clusters $\mathcal{C} = \{{c}_1, \dots, {c}_s \}$, as shown in \figurename~\ref{fig:bird}. The segmentation can be a result of either the geographical distance or the specific operational requirements of different entities that only utilise certain parts of the IoT network. An example of such a large-scale system comprising massive IoT network and distributed processing across a large deployment of 5G and beyond-5G radio access network and edge computing nodes are various smart grid applications (see \cite{cosovic20175g} and references therein). Emerging applications also include a number of 5G and beyond-5G distributed network functions such as device localisation, handover management, radio resource and energy management and many others~\cite{loven2019edgeai}.

Clustering allows running GBP in distributed fashion on a multi-core computing platform in a single physical node, or to run subgraphs at different and physically separated physical nodes, mutually inter-connected via a communication network. In the following, we consider the latter, and adopt the GBP message scheduling tailored to the distributed architecture. We proceed by outlining the methodology for graph clustering and distributed message scheduling, taking into account communication constraints between clusters.

\subsection{Factor Graph Clustering} 
\begin{figure}[ht]
	\centering
	\includegraphics[width=7.9cm]{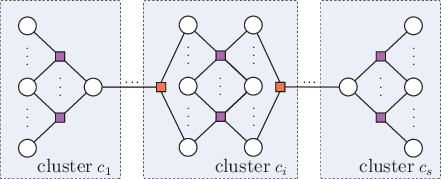}
	\caption{The factor graph allocated to different clusters with internal factor nodes (purple squares) and tie factor nodes (orange squares).}
	\label{fig:clusters}
\end{figure}

Factor graph representation allows us to flexibly distribute architecture given in \figurename~\ref{fig:bird} and defined by \eqref{eqn:linear_model_matrix} over the set of clusters $\mathcal{C}$. Each IoT device gathers and observes particular data of interest. This data is a function of state variables positioned in a close proximity of IoT device. The goal of the inference process over the factor graph is to estimate the values of these state variables. In the factor graph representation, each IoT device is associated with a factor node, and each state variable with a variable node. Hence, each cluster ${c}_i$ is a connected subgraph induced from a disjoint subset of variable nodes of the original factor graph $\mathcal{X}_{c_i} \subset \mathcal{X}$, as illustrated in \figurename~\ref{fig:clusters}. As a result, we obtain the set of \emph{internal} factor nodes $\mathcal{I} \subset \mathcal{F}$, and the set of \emph{tie} factor nodes $\mathcal{T} \subset \mathcal{B}$, where $\mathcal{F} = \{ \mathcal{I}, \mathcal{T} \}$. Internal factor nodes from the set $\mathcal{I}_{c_i} \subset \mathcal{I}$, are only associated with the cluster ${c}_i$, connecting exclusively variable nodes from the set $\mathcal{X}_{c_i}$. In contrast, tie factor nodes from the set $\mathcal{T}_{c_i} \subset \mathcal{T}$ connect variable nodes from the set $\mathcal{X}_{c_i}$ to the variable nodes from the set $\mathcal{X}_{c_j}$, $\forall c_j$, that belong to clusters $c_j \in \mathcal{C} \setminus c_i$.

In the distributed framework, the linear models \eqref{eqn:linear_model_matrix} can be represented as follows: 
\begin{equation}
    \resizebox{0.89\hsize}{!}{%
    $\begin{bmatrix}
        \mathbf{H}_{c_1} & \mathbf{H}_{c_1,c_2} & \dots & \mathbf{H}_{c_1,c_s} \\
        \mathbf{H}_{c_2,c_1} & \mathbf{H}_{c_2} & \dots & \mathbf{H}_{c_2,c_s} \\
        \vdotswithin{} \\
        \mathbf{H}_{c_s,c_1} & \mathbf{H}_{c_s,c_2} & \dots & \mathbf{H}_{c_s}
    \end{bmatrix}
    \begin{bmatrix}
        \mathbf{x}_{c_1} \\
        \mathbf{x}_{c_2} \\
        \vdotswithin{} \\
        \mathbf{x}_{c_s}
    \end{bmatrix} +
    \begin{bmatrix}
        \mathbf{u}_{c_1} \\
        \mathbf{u}_{c_2} \\
        \vdotswithin{} \\
        \mathbf{u}_{c_s}
    \end{bmatrix} =
    \begin{bmatrix}
        \mathbf{z}_{c_1} \\
        \mathbf{z}_{c_2} \\
        \vdotswithin{} \\
        \mathbf{z}_{c_s}
    \end{bmatrix},$
    }
    \label{eqn:cluster_matrix}
\end{equation}
where each cluster $c_i$ is defined by the \emph{internal} matrix $\mathbf{H}_{c_i} \in \mathbb{R}^{m_{c_i} \times n_{c_i}}$ and \emph{tie} matrices $\mathbf{H}_{c_i, c_j} \in \mathbb{R}^{m_{c_i} \times n_{c_j}}$, $\forall c_j$, $c_j \in \mathcal{C} \setminus c_i$. The matrix $\mathbf{H}_{c_i}$ encompasses variable nodes $\mathcal{X}_{c_i}$, while the matrix $\mathbf{H}_{c_i, c_j}$ ties the cluster $c_j \in \mathcal{C} \setminus c_i$ with the set of variable nodes $\mathcal{X}_{c_j}$. To define a factor node $f_i$, let us take an arbitrary row $\mathbf{h}_{c_i}$ of the matrix $\mathbf{H}_{c_i}$, whose at least one element is nonzero. For the selected row $\mathbf{h}_{c_i}$, we take corresponding rows $\mathbf{h}_{c_i, c_j}$ across the matrices $\mathbf{H}_{c_i, c_j}$, $c_j \in \mathcal{C} \setminus c_i$. A factor node $f_i$ is defined by conjunction of rows formed by $\mathbf{h}_{c_i}$ and $\mathbf{h}_{c_i, c_j}$, $\forall c_j$. For the case where the all elements of $\mathbf{h}_{c_i, c_j}$, $\forall c_j$, $c_j \in \mathcal{C} \setminus c_i$ are equal to zero, the factor node connections are defined by nonzero elements of $\mathbf{h}_{c_i}$, forming internal factor node $f_i \in \mathcal{I}_{c_i}$. In contrast, nonzero elements of $\mathbf{h}_{c_i, c_j}$ represent connections that tie (i.e., connect) variable nodes belonging to the cluster $c_j \in \mathcal{C} \setminus c_i$, thus defining the tie factor node $f_i \in \mathcal{T}_{c_i}$

We focus on the case where a large factor graph is segmented into clusters with the vast majority of factor nodes representing internal factor nodes, and with only a small fraction of tie factor nodes. In other words, for each cluster $c_i$, the number of non-zero elements $\lambda_{c_i}$ of the internal matrix $\mathbf{H}_{c_i}$ is significantly larger than the number of non-zero elements $\gamma_{c_i}$ of the tie matrices $\mathbf{H}_{c_i, c_j}$, $\forall c_j$, $c_j \in \mathcal{C} \setminus c_i$. Note that $\lambda_{c_i}$ and $\gamma_{c_i}$ define a number of internal and tie edges of the subgraph belonging to the cluster $c_i$.

\subsection{Message Scheduling in Distributed Framework}
The main drawback of the synchronous GBP algorithm when applied to the distributed scenario is potentially high communication delay over the tie factor nodes $\mathcal{T}$ compared to the time needed to complete a single iteration within a cluster over the internal factor nodes $\mathcal{I}$. Let us start by observing the segmented factor graph without tie factor nodes $\mathcal{T}$. In each cluster $c_i \in \mathcal{C}$, let us assume that the processing delay of variable-node and factor-node processing is $\tau_x^{(c_i)}$ and $\tau_f^{(c_i)}$, respectively. After connecting clusters using tie factor nodes $\mathcal{T}_{c_i}$, the duration of processing causes an additional time delay $\tau_c^{(c_i)}$. This inter-cluster delay may be due to exchange of messages between the cores in a multi-core processing node or exchange of messages between the nodes in a communication network. Thus, the synchronous GBP on the segmented factor graph produces the estimate after $\nu$ iterations at the time instant $\nu(\tau^{(c_i)}_{m} + \tau_c)$, under assumption that the additional delay $\tau_c^{(c_i)}$ is constant and independent of the clusters $\tau_c = \tau_c^{(c_i)}$, and where $\tau^{(c_i)}_{m} = \max \{\tau^{(c_i)}_{x} + \tau^{(c_i)}_{f}: i = 1,\dots,s \}$. In comparison with the synchronous GBP applied over the complete factor graph, the message processing distributed across multiple clusters is less time consuming $\tau^{(c_i)}_{m} < \tau_{m}$ (due to parallelism), however, the duration of the delay $\tau_c > \tau_m$ over tie factor nodes $\mathcal{T}$ may lead to a significant overall time delay in the calculation of the estimate $\hat{\mathbf{x}}^{(\nu)}$.

To minimise delay over tie factor nodes, we consider different message scheduling strategies on the segmented factor graph aiming to produce an accurate estimate $\hat{\mathbf{x}}$ with minimal time delay in static and dynamic scenarios. The objective of distributed message passing scheduling is to reduce the duration $\nu\tau_c$ while maintaining accuracy and convergence. We propose an \emph{alternating scheduling} based on sequences of the exchanging messages using global (inter-cluster) and local (intra-cluster) iterations. Each sequence $k_s = \{1,2,\dots \}$ consists of global $k_g = \{1,\dots, \nu_g\}$ and local iterations $k_l = \{1,\dots, \nu_l\}$ of the GBP algorithm. Intuitively, condition $\nu_l \geq \nu_g$ is preferred for the AGBP algorithm, under conditions $\lambda_{c_i} > \gamma_{c_i}$ and $\tau_c > \tau_m$.         

\begin{figure}[ht]
	\centering
	\includegraphics[width=7.9cm]{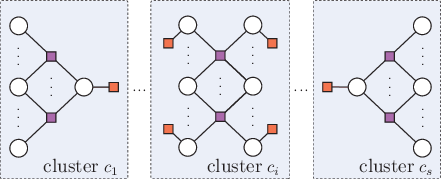}
	\caption{The disjoint factor graph allocated to different clusters, where tie factor nodes are collapsed into single connected factor nodes.}
	\label{fig:clusters_local}
\end{figure}

At the level of global or inter-cluster iterations, we compute all messages $\bm{\mu}_{x} = [{
\mu}_{x_j \to f_i}]$ and $\bm{\mu}_{f} = [\mu_{f_i \to x_j}]$, $f_i \in \mathcal{B}$, $x_j \in \mathcal{X}$, corresponding to synchronous scheduling over the complete factor graph, as shown in \figurename~\ref{fig:clusters}. At the level of local or intra-cluster iterations, we perform local iterations for each cluster by calculating the messages over disjoint segments of the factor graph, as shown in \figurename~\ref{fig:clusters_local}. 
In particular, we compute messages $\bm{\mu}_{x} = [{\mu}_{x_j \to f_i}]$ and $\bm{\mu}_{f} = [\mu_{f_i \to x_j}]$, $f_i \in \mathcal{B} \setminus \mathcal{T}$, $x_j \in \mathcal{X}$, where each tie factor node $f_i \in \mathcal{T}$ is collapsed into $\text{deg}(f_i)$ leaf factor nodes. The messages from these leaf factor nodes are equal to $\bm{\mu}_{f} = [\mu_{f_i \to x_j}]$, $f_i \in \mathcal{T}$, $x_j \in \mathcal{X}$ obtained after the last global iteration.\footnote{In Appendix B, we have included a detailed description of the AGBP algorithm using a practical example to guide the reader through each step.} 

Combining global and local iterations, the AGBP algorithm reaches a given convergence condition after $\nu_s(\nu_g + \nu_l)$ iterations, where $k_s=\nu_s$ is the total number of the sequences. As a result, the AGBP yields the estimate $\hat{\mathbf{x}}^{(\nu_s(\nu_g + \nu_l))}$ at the time instant $\nu_s\nu_g(\tau^{(c_i)}_{m} + \tau_c) + \nu_s\nu_l\tau^{(c_i)}_{m}$. Let us assume $\tau^{(c_i)}_{m} \ll \tau_c$, implying $\nu_s\nu_l\tau^{(c_i)}_{m} \ll \nu_s\nu_g(\tau^{(c_i)}_{m} + \tau_c)$. The AGBP may be less time consuming than synchronous GBP algorithm $\nu_s\nu_g \tau_c < \nu\tau_m$ only if $\nu_s\nu_g < \nu$, where $\tau_c > \tau_m$. As demonstrated in the numerical section, the condition $\nu_s\nu_g < \nu$ is valid, in addition to observation that $\nu_s(\nu_g + \nu_l) < \nu$ also holds across different scenarios. This inequality implies that AGBP seeds the more rapid convergence rate\footnote{The convergence rate defines the rate (i.e., measure) at which the distance $||\hat{\mathbf{x}}^{(k)} - \hat{\mathbf{x}}|| \to 0$ as $k \to \infty$~\cite[Sec.~4.6]{saleri}.}. In the following subsection, we outline how the scheduling of the messages in conjunction with the properties of the linear model \eqref{eqn:linear_model_matrix} affects the convergence rate.

\subsection{Convergence Analysis of Alternating Gaussian Belief Propagation Algorithm}
We present a convergence analysis of the AGBP algorithm. In general, the convergence of the GBP correlates with the spectral radius of the matrix that affects the evolution of the mean values $\mathbf{m}_{f} = [m_{f_i \to x_j}]$ of the messages $\bm{\mu}_{f}= [\mu_{f_i \to x_j}(x_j)]$, $f_i \in \mathcal{B}$, $x_j \in \mathcal{X}$. We note that the corresponding variances $\mathbf{v}_{f} = [v_{f_i \to x_j}]$ always converge to the finite-valued fixed point $\mathbf{v}_{f}^* = [v_{f_i \to x_j}^*]$ \cite{cosovictra, Azhang}. Without loss of generality, let us observe the factor graph with $b = |\mathcal{B}|$ branch factor nodes from the set $\mathcal{B} = \{f_1, \dots, f_b\}$, $b < m$, and $l = |\mathcal{L}|$ leaf factor nodes defined by the set $\mathcal{L} = \{f_{b+1}, \dots, f_m\}$. To recall, each row of the matrix $\mathbf{H}$ corresponds to one factor node, therefore, the first $b$ rows of the matrix $\mathbf{H}$ correspond to branch factor nodes, while leaf factor nodes start from the row $b+1$. 

Let us start with the convergence analysis of the synchronous GBP algorithm. Using \eqref{BP_vf_mean} and \eqref{BP_fv_mean}, the evolution of the mean values can be written as follows:  
\begin{equation}
    \mathbf{m}^{(k)}_{f} = \mathbf{c}_f + \bm {\Omega} \mathbf{m}^{(k-1)}_{f}, 
    \label{eqn:synchronous_means}    
\end{equation}
where the vector $\mathbf{m}_{f} = [\mathbf{m}_{f_1}, \dots, \mathbf{m}_{f_b}]^T \in \mathbb{R}^{d}$, $d = \sum_{f_i \in \mathcal{B}} \text{deg}(f_i)$, represents the mean values of messages from branch factor nodes $\mathcal{B}$ to variable nodes $\mathcal{X}$, with the $i$-th vector defined as $\mathbf{m}_{f_i} = [m_{f_i \to x_j}]$, $f_i \in \mathcal{B}$, $x_j \in \mathcal{X}_{i}$. The vector $\mathbf{c}_{f} = [\mathbf{c}_{f_1}, \dots, \mathbf{c}_{f_b}]^T \in \mathbb{R}^d$ represents the constant component of \eqref{eqn:synchronous_means}. The matrix $\bm {\Omega} = [\bm{\Omega}_{f_{1}},\dots,\bm{\Omega}_{f_{b}}]^T \in \mathbb{R}^{d \times d}$ affects the evolution of the mean values $\mathbf{m}_{f}$, where the $i$-th block $\bm{\Omega}_{f_i} \in \mathbb{R}^{d_i \times d}$, $d_i = \text{deg}(f_i)$. The definition of matrix $\bm {\Omega}$ and vector $\mathbf{c}_f$, involved in \eqref{eqn:synchronous_means}, is provided in Appendix C.

\begin{theorem}
The means $\mathbf{m}_{f}$ converge to a unique fixed point $\lim_{k \to \infty}\mathbf{m}_{f}^{(k)} = \mathbf{m}_f^{*}$:
\begin{equation}
    \mathbf{m}_f^{*} = (\mathbf{I} - \bm{\Omega})^{-1}\mathbf{c}_f,
    \label{eqn:fixed_point}
\end{equation} 
for any initial point $\mathbf{m}_{f}^{(0)}$ if and only if the spectral radius $\rho(\bm {\Omega}) < 1$.
\end{theorem}
\begin{proof}\renewcommand{\qedsymbol}{}
The proof follows similar steps as in Theorem 5.2 in \cite{hanly}.
\end{proof}

Global iterations are executed over the entire factor graph and equate the convergence analyses of the synchronous GBP algorithm. According to \eqref{eqn:synchronous_means}, the evolution equation can be written as follows:
\begin{equation}
    \mathbf{m}^{(k_s, k_g)}_{f} = \mathbf{c}_f + \bm {\Omega} \mathbf{m}^{(k_s, k_g-1)}_{f},
    \label{eqn:recursion_global}    
\end{equation} 
where the initial point $\mathbf{m}^{(k_s,0)}_{f}$ is equal to $\mathbf{m}^{(k_s-1,\nu_l)}_{f}$ or $\mathbf{m}^{(k_s,\nu_l)}_{f}$ depending if the corresponding sequence $k_s$ is initialised with global or local iterations.

We complement the previous analysis with the case that includes local iterations, where each tie factor node $f_i \in \mathcal{T} \subset \mathcal{B}$ transforms to $\text{deg}(f_i)$ leaf factor nodes. The resulting factor graph contains $b-g$, $g = |\mathcal{T}|$, branch factor nodes and $l + e$, $e = \sum_{f_i \in \mathcal{T}} \text{deg}(f_i)$, leaf factor nodes. Hence, local iterations operate on the altered system \eqref{eqn:recursion_global}, whose dimensionality is reduced by $e$ dimensions. This reduction of space can be achieved simply by replacing nonzero elements with zeros for each block $\bm{\Omega}_{f_{i}}$ , $f_i \in \mathcal{T}$, of the matrix $\bm {\Omega}$. Then, the evolution of means can be written as follows:     
\begin{equation}
    \mathbf{m}^{(k_s, k_l)}_{f} = \mathbf{Q} {\mathbf{c}}_f + (\mathbf{I} - \mathbf{Q}) \mathbf{m}^{(k_s,\nu_g)}_{f} + \mathbf{Q}\bm {\Omega} \mathbf{m}^{(k_s,k_l-1)}_{f}.  
    \label{eqn:recursion_means_local}    
\end{equation}  
Similarly to global iterations, the initial point $\mathbf{m}^{(k_s,0)}_{f}$ depends on the schedule of global and local iterations in the corresponding sequence $k_s$, and can be $\mathbf{m}^{(k_s,\nu_g)}_{f}$ or $\mathbf{m}^{(k_s-1,\nu_g)}_{f}$. The matrix $\mathbf{Q} = \text{diag}(\mathbf{Q}_{f_1}, \dots, \mathbf{Q}_{f_b})\in \mathbb{F}^{d \times d}$, $\mathbb{F}=\{0, 1\}$ is a block-diagonal, and $\mathbf{I}$ is $d \times d$ identity matrix. The $i$-th block of the matrix $\mathbf{Q}$ is equal to $\mathbf{Q}_{f_i} = \mathbf{I}_{f_i}$, $f_i \in \mathcal{B} \setminus \mathcal{T}$, or $\mathbf{Q}_{f_i} = \mathbf{0}_{f_i}$, $f_i \in \mathcal{T}$, where $\mathbf{I}_{f_i}$ is $d_i \times d_i$ identity matrix, and $\mathbf{0}_{f_i}$ is a matrix $d_i \times d_i$ of zeros. The resulting vector $\mathbf{Q}\mathbf{c}_f$ consists only of mean values from $b-g$ branch factor nodes, while the vector $(\mathbf{I} - \mathbf{Q}) \mathbf{m}_f^{(k_s,\nu_g)}$ contains mean values from $g$ branch factor nodes obtained in the last global iteration. 

The AGBP algorithm sequentially alternates equations \eqref{eqn:recursion_global} and \eqref{eqn:recursion_means_local}, arriving to the solution representing exact means only if algorithm converges. The following theorem formalises this observation.
\begin{theorem}
Let $\mathbf{m}_f^{*}$ be a fixed point of the synchronous GBP, then if the AGBP algorithm converges, the resulting fixed point is equal to $\mathbf{m}_{f}^{*}$. 
\end{theorem}
\begin{proof}\renewcommand{\qedsymbol}{}
The proof can be found in the Appendix C.
\end{proof}

We note that characterising the convergence rate is a challenging task even in the case of synchronous GBP \cite{zhang2020convergence}, thus we leave it out the scope of this paper. 

\subsection{Dynamic Gaussian Belief Propagation Algorithm}
From \eqref{eqn:synchronous_means}, as well as \eqref{eqn:recursion_global} and \eqref{eqn:recursion_means_local}, it follows that the initial values of the factor node to the variable node messages affect the number of iterations needed for convergence. In particular, the large number of iterations can be caused by random message initialisation, in cases when priors of their means and variances do not exist. In dynamic scenarios, where new data arrive and evolve over time, it is straightforward to  consider GBP as an efficient continuous inference algorithm. In this scenario, upon arrival of new data, all messages are close to their fixed-point values, and initialising GBP from this point allows it to converge in fewer iterations. 

We develop a framework for analysing dynamic systems by simulating arrival of new observations, which are further infused with deterioration or ageing component over time. We integrate these data into the continuously running instances of the GBP algorithm. The proposed ageing solution relies on the GBP algorithm robustness to the ill-conditioned scenarios caused by significant differences between values of variances. This overcomes the main drawback of traditional solutions based on the weighted least-squares or recursive least-squares algorithms. More precisely, the GBP is capable of integrating a wide range of variances, from small values $v_i \to 0$ to large values $v_i \to \infty$~ \cite{cosovicfast}. This property allows inference over the factor graph that reflects the entire network of sensors, regardless of whether the data is available at a given time. For example, data arriving at an arbitrary time instant affects the behaviour of the system according to its finite variance $0<v_i<\infty$, while data without impact are represented in the factor graph via infinite variance $v_i \to \infty$. These features are key foundations for the dynamic GBP algorithm, where the variance values of different factor nodes change dynamically over time.

Let $\tau_r$, $r=1,2,\dots$, denote time instants when the factor node $f_i$ receives the observation value $z_i$ with the predefined initial variance $v_{i}$. After each time instant $\tau_r$, the dynamic model increases the variance value $v_{i}$ over time. At each time interval $\tau_r \leq t < \tau_{r+1}$, we associate the Gaussian distribution $\mathcal{N}(z_i(t)|\mathcal{X}_i,v_i(t))$ with the corresponding factor node $f_i$, where the variance $v_i(t)$ increases its value starting from the predefined variance $v_i$, while the mean value $z_i(t)$ remains constant according to the received value $z_i$. Depending on the dynamic arrival of data, an adaptive mechanism for increasing the variance over the time $v_i(t)$ can be defined. The logarithmic growth model represents a promising solution for systems with a high sampling rate of the data, where a rapid increase in variance is required: 
\begin{equation}
    v_i(t) = \alpha \, \text{log} \left(\frac{t-\tau_r + 1+\beta}{1+\beta} \right ) + v_i, \;\ \tau_r \leq t < \tau_{r+1},
    \label{growth_log}    
\end{equation}
where parameters $\alpha$ and $\beta$ controls the rate of the growth. In contrast, the exponential growth model corresponds to systems with a low sampling rate of the data:
\begin{equation}
    v_i(t) = v_i(1+\beta)^{\alpha(t-\tau_r)}, \;\ \tau_r \leq t < \tau_{r+1}.
    \label{growth_exp}
\end{equation}
Finally, the linear growth model can be observed as a compromise between logarithmic and exponential growth models:
\begin{equation}
    v_i(t) = \alpha(t-\tau_r) + v_i, \;\ \tau_r \leq t < \tau_{r+1}.
    \label{growth_lin}    
\end{equation}

\begin{figure}[ht]
	\centering
	\captionsetup[subfigure]{oneside,margin={0.9cm,0cm}}
	\begin{tabular}{@{\hspace{-0.4cm}}c@{}}
	\subfloat[]{\label{fig_log}
	\begin{tikzpicture}
	    \pgfmathsetmacro{\Ttau}{4}; \pgfmathsetmacro{\Trho}{12}; \pgfmathsetmacro{\Ttheta}{20}; 
        \pgfmathsetmacro{\TtauN}{28}; \pgfmathsetmacro{\TrhoN}{36}; \pgfmathsetmacro{\Vr}{1}
        \begin{axis}[width=3.55cm,height=4.0cm, grid=both, grid style={line width=.3pt, draw=gray!60},
            label style={font=\footnotesize}, ticklabel style = {font=\footnotesize}, ylabel shift = -5 pt,
            xlabel={$t$}, ylabel={$v_i(t)$},
            xmin = 0, xmax = 40, ymin = 0.5, ymax = 4,
            xtick={4,8,12,16,20,24,28,32,36}, xticklabels={$\tau_1$,,$\rho_1$,,$\theta_1$,,$\tau_{2}$,,$\rho_2$},  
            ytick={1,2.33,3.66}, yticklabels={$v_i$,,$\bar{v}_i$},
            every axis plot/.append style={thick}, font=\footnotesize,
            legend columns=3,  legend style={/tikz/column 2/.style={column sep=6pt,},},
            legend style={/tikz/column 4/.style={column sep=6pt,},}]
            \addplot [black, domain=0:\Ttau, samples = 20] {3.66};
            \addplot [black, samples = 20] coordinates {(\Ttau,3.66)(\Ttau,\Vr)};
            \addplot [black, domain=\Ttau:\Trho, samples = 20] {\Vr};
            \addplot [black, domain=\Trho:\Ttheta, samples = 20] {3 * (log10((x - \Trho + 1.2)/1.2)) + \Vr};
            \addplot [black, domain=\Ttheta:\TtauN, samples = 20] {3 * (log10((\Ttheta - \Trho + 1.2)/1.2)) + \Vr};
            \addplot [black, samples = 20] coordinates {(\TtauN,3.66)(\TtauN,\Vr)};
            \addplot [black, domain=\TtauN:\TrhoN, samples = 20] {\Vr};
            \addplot [black, domain=\TrhoN:40, samples = 20] {3 * (log10((x - \TrhoN + 1.2)/1.2)) + \Vr};
        \end{axis}
    \end{tikzpicture}}
	\end{tabular}
	\captionsetup[subfigure]{oneside,margin={1.2cm,0.6cm}}
	\begin{tabular}{@{}c@{}}
	\subfloat[]{\label{fig_exp}
    \begin{tikzpicture}
	    \pgfmathsetmacro{\Ttau}{4}; \pgfmathsetmacro{\Trho}{12}; \pgfmathsetmacro{\Ttheta}{20}; 
        \pgfmathsetmacro{\TtauN}{28}; \pgfmathsetmacro{\TrhoN}{36}; \pgfmathsetmacro{\Vr}{1}
        \begin{axis}[width=3.55cm,height=4.0cm, grid=both, grid style={line width=.3pt, draw=gray!60},
            label style={font=\footnotesize}, ticklabel style = {font=\footnotesize}, ylabel shift = -5 pt,
            xlabel={$t$}, 
            xmin = 0, xmax = 40, ymin = 0.5, ymax = 4.7,
            xtick={4,8,12,16,20,24,28,32,36}, xticklabels={$\tau_1$,,$\rho_1$,,$\theta_1$,,$\tau_{2}$,,$\rho_2$},  
            ytick={1,2.64,4.28}, yticklabels={$v_i$,,$\bar{v}_i$},
            every axis plot/.append style={thick}, font=\footnotesize,
            legend columns=3,  legend style={/tikz/column 2/.style={column sep=6pt,},},
            legend style={/tikz/column 4/.style={column sep=6pt,},}]
            \addplot [black, domain=0:\Ttau, samples = 20] {4.28};
            \addplot [black, samples = 20] coordinates {(\Ttau,4.28)(\Ttau,\Vr)};            
            \addplot [black, domain=\Ttau:\Trho, samples = 20] {\Vr};
            \addplot [black, domain=\Trho:\Ttheta, samples = 20] {\Vr*((1+0.199)^(x-\Trho))};
            \addplot [black, domain=\Ttheta:\TtauN, samples = 20] {\Vr*((1+0.199)^(\Ttheta-\Trho))};
            \addplot [black, samples = 20] coordinates {(\TtauN,4.28)(\TtauN,\Vr)};
            \addplot [black, domain=\TtauN:\TrhoN, samples = 20] {\Vr};
            \addplot [black, domain=\TrhoN:40, samples = 20] {\Vr*((1+0.199)^(x-\TrhoN))};
        \end{axis}
    \end{tikzpicture}}
	\end{tabular}
	\begin{tabular}{@{}c@{}}
	\subfloat[]{\label{fig_lin}
	\begin{tikzpicture}
	    \pgfmathsetmacro{\Ttau}{4}; \pgfmathsetmacro{\Trho}{12}; \pgfmathsetmacro{\Ttheta}{20}; 
        \pgfmathsetmacro{\TtauN}{28}; \pgfmathsetmacro{\TrhoN}{36}; \pgfmathsetmacro{\Vr}{1}
        \begin{axis}[width=3.55cm,height=4.0cm, grid=both, grid style={line width=.3pt, draw=gray!60},
            label style={font=\footnotesize}, ticklabel style = {font=\footnotesize}, ylabel shift = -5 pt,
            xlabel={$t$},
            xmin = 0, xmax = 40, ymin = 0.5, ymax = 4.7,
            xtick={4,8,12,16,20,24,28,32,36}, xticklabels={$\tau_1$,,$\rho_1$,,$\theta_1$,,$\tau_{2}$,,$\rho_2$}, 
            ytick={1,2.64,4.28}, yticklabels={$v_i$,,$\bar{v}_i$},
            every axis plot/.append style={thick}, font=\footnotesize,
            legend columns=3,  legend style={/tikz/column 2/.style={column sep=6pt,},},
            legend style={/tikz/column 4/.style={column sep=6pt,},}]
            \addplot [black, domain=0:\Ttau, samples = 20] {4.28};
            \addplot [black, samples = 20] coordinates {(\Ttau,4.28)(\Ttau,\Vr)}; 
            \addplot [black, domain=\Ttau:\Trho, samples = 20] {\Vr};
            \addplot [black, domain=\Trho:\Ttheta, samples = 20] {0.412*(x-\Trho)+\Vr};
            \addplot [black, domain=\Ttheta:\TtauN, samples = 20] {\Vr*((1+0.199)^(\Ttheta-\Trho))};
            \addplot [black, samples = 20] coordinates {(\TtauN,4.28)(\TtauN,\Vr)};
            \addplot [black, domain=\TtauN:\TrhoN, samples = 20] {\Vr};
            \addplot [black, domain=\TrhoN:40, samples = 20] {0.412*(x-\TrhoN)+\Vr};
        \end{axis}
    \end{tikzpicture}} 
	\end{tabular}
	\caption{The logarithmic growth model (subfigure a), the exponential growth model (subfigure b), and the linear growth model (subfigure c).}
	\label{fig_aging}%
\end{figure}
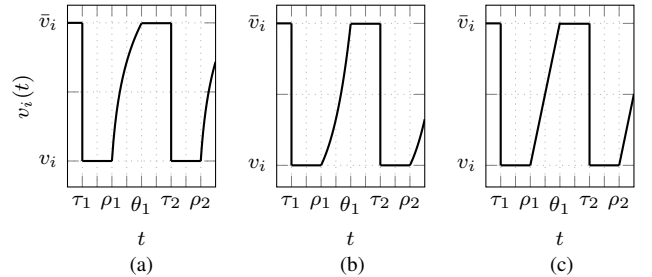

In addition, we can fix the predefined variance $v_i$ as a constant value for a certain period of time $\tau_r \leq t \leq \rho_r$, which is especially advantageous in networks whose dynamics change slowly. In practise, the dynamic model requires defining a limit from above $\bar {v}_i$ of a function $v_i(t)$, instead of allowing variance to take on extremely large values, especially if the broadcast GBP algorithm is used \cite{bickson}. The variance will increase the value over the period $\rho_r \leq t \leq \theta_r$, after which it keeps the constant value $\bar {v}_i = v_i(\theta_r)$ over the time period $\theta_r \leq t < \tau_{r+1}$. To summarise, when the computation unit receives the data at the time instant $\tau_r$, the value of variance remains constant $v_i$ up to the time instant $\rho_r$, followed by the variance increase up to the time instant $\theta_r$, when its value approaches saturation $\bar {v}_i$. An extension of the growth models \eqref{growth_log} - \eqref{growth_lin} defined in this way allows for more flexible ageing approach, shown in~\figurename\ref{fig_aging}. As an example, the comprehensive logarithmic growth model can be written in the following form:
\begin{equation}
    \resizebox{0.89\hsize}{!}{%
    $v_i(t) = \begin{cases} 
      v_i, & \tau_r \leq t \leq \rho_r   \\
      \alpha \, \text{log} \left(\cfrac{t-\rho_r + 1+\beta}{1+\beta} \right ) + v_i, & \rho_r \leq t\leq \theta_r  \\
      \bar {v}_i, & \theta_r \leq t < \tau_{r+1}.
  \end{cases}$
  }
  \label{logarithmic}
\end{equation}

To conclude, the presented approach allows the inclusion of an arbitrary number of sensors with extremely large variance values with negligible impact on the system. Parts of the system obtain a unique solution based on real-time data, while the rest of the system will be determined based on both real-time data and data with large variance values.  

\section{Numerical Results}
In this section, we analyse distributed linear models in the form defined by \eqref{eqn:cluster_matrix}. We artificially create matrices with independent and identically distributed (i.i.d.) nonzero entries distributed uniformly in the interval $[0, 1)$. To be more precise, for each cluster $c_i$, we form internal matrix $\mathbf{H}_{c_i}$ and tie matrices $\mathbf{H}_{c_i,c_j}$, $\forall c_j$, $c_j \in \mathcal{C}\setminus c_i$, according to the internal and tie sparsity. Sparsity governs the expected number of the factor graph edges, denoting it $\bar{\lambda}_{c_i}$ and $\bar{\gamma}_{c_i}$ for internal and tie edges, respectively. For each simulated scenario, characterised by a set of distributed linear model's parameters such as matrix dimensions, expected number of internal and tie edges, number of clusters, etc., simulations are repeated 500 times in order to get statistically significant results. 

Due to inherent differences in procedures for calculation of the solution $\hat{\mathbf{x}}$ between synchronous and AGBP, we consider the following identity $\nu\tau_m =$ $\nu_s\nu_g(\tau^{(c_i)}_{m} + \tau_c) +$ $\nu_s\nu_l\tau^{(c_i)}_{m}$ that equates the convergence time of both methods. We proceed by comparing the processing delays of two GBP versions as follows. Using this identity, we express the inter-cluster delay $\tau_c$ as follows:
\begin{equation}
    \tau_c = \frac{\nu - \kappa\nu_s(\nu_g + \nu_l)}{\nu_s\nu_g}\tau_m = \phi \tau_m ,
    \label{eqn:metric_time}
\end{equation}
where under the condition $\lambda_{c_i} > \gamma_{c_i}$, processing time $\tau^{(c_i)}_{m}$ is estimated as $\kappa \tau_{m}$, $\kappa = \max\{ \lambda_{c_i}/ \sum_{j=1}^s (\lambda_{c_j} + \gamma_{c_j}): \forall c_i \}$, $0<\kappa<1$. The constant $\phi$ quantifies the critical  value of a ratio between the inter-cluster message delay $\tau_c$ and the duration of a single iteration of synchronous GBP $\tau_m$, under which both GBP algorithms converge simultaneously. In other words, for all the values of $\tau_c < \phi \tau_m$, the AGBP algorithm converges faster than the synchronous GBP. In order to obtain the number of iterations $\nu$ and the number of sequences $\nu_s$ needed for convergence of synchronous GBP and AGBP, respectively, we run both algorithms until the root mean square error between the maximum likelihood solution of \eqref{eqn:linear_model_matrix} and the GBP estimate reaches $10^{-5}$. 

\subsection{Linear Models with Symmetric Matrices Properties} \label{symmetric}
We start our experiments by analysing performance of GBP in an environment with well-defined matrix properties. We observe distributed linear models, where each cluster consists of the symmetric matrix $\mathbf{H}_{c_i}$, and matrices $\mathbf{H}_{c_i,c_j} = \mathbf{H}_{c_j,c_i}^T$, $m_{c_i} = n_{c_i} = n_{c_j}$. This setup forms the symmetric matrix $\mathbf{H} \in \mathbb{R}^{m \times n}$, $m=n$, where diagonal elements are calculated according to $h_{ii} = \sum_{i\neq j}h_{ij} + \delta$, $i=1,\dots,n$. Note that a square matrix $\mathbf{H}$ has full rank, resulting in a solution independent of observation variances. 

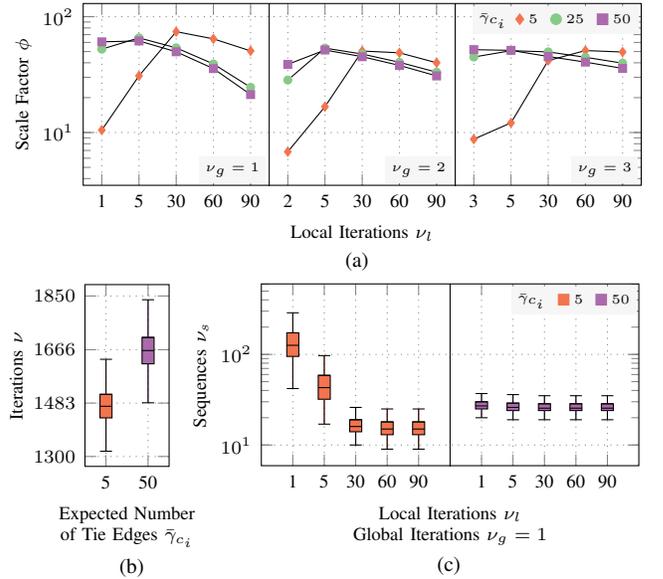
\begin{figure}[ht]
	\centering \captionsetup[subfigure]{oneside,margin={1.0cm,0cm}}
	\begin{tabular}{@{\hspace{-0.2cm}}c@{}} 
	\subfloat[]{\label{plot1a} \centering	
        \begin{tikzpicture}
  	        \begin{axis}[width=9cm, height=4.0cm, ymode=log,
	            grid style={line width=.3pt, draw=gray},
                x tick label style={/pgf/number format/.cd, set thousands separator={},fixed},
                ylabel style ={yshift=-0.1cm},
  	            xlabel={Local Iterations $\nu_l$},
  	            ylabel={Scale Factor $\phi$},
  	            label style={font=\scriptsize},
  	            grid=major,
  	            ymin = 3.5, ymax = 130, xmin = 0.5, xmax = 15.5,
  	            xtick={1,2,3,4,5, 6,7,8,9,10, 11,12,13,14,15},
  	            xticklabels={1,5,30,60,90, 2,5,30,60,90, 3,5,30,60,90},
  	            tick label style={font=\scriptsize},
  	            legend style={draw=none, fill=gray!7!white, legend cell align=left, font=\tiny,
  	            at={(0.995, 0.985)}, anchor=north east, 
  	            {/tikz/every even column/.append style={column sep=0.1cm}}, 
  	            {/tikz/column 1/.append style={column sep=-0.2cm}},
  	            /tikz/column 4/.style={yshift=-0.025cm}, 
  	            /tikz/column 6/.style={yshift=-0.025cm}, 
  	            /tikz/column 8/.style={yshift=-0.025cm}, inner sep=0.5pt}, 
  	            legend columns=4]
                
                \addlegendimage{legend image with text=$\bar{\gamma}_{c_i}$}
                \addlegendentry{}; 
	            \addlegendimage{mark=diamond*, only marks, mark options={wind}}
	            \addlegendentry{$5$}; 
                \addlegendimage{mark=otimes*, only marks, mark options={gas}}
	            \addlegendentry{$25$}; 
	            \addlegendimage{mark=square*, only marks, mark options={biofuels}}
	            \addlegendentry{$50$}; 

                \addplot[mark=diamond*, mark size=1.7pt, mark options={wind}, black] 
  	            table [x={x}, y={y}] {./plot/plot1/plot1a/area006_tie5_global1.txt}; 
	            \addplot[mark=otimes*, mark size=1.5pt, mark options={gas}, black] 
  	            table [x={x}, y={y}] {./plot/plot1/plot1a/area006_tie25_global1.txt}; 
  	            \addplot[mark=square*, mark size=1.5pt, mark options={biofuels}, black]
  	            table [x={x}, y={y}] {./plot/plot1/plot1a/area006_tie50_global1.txt}; 
  	            \draw [thin] (axis cs:5.5,0.1) -- (axis cs:5.5,180);
  	            \node[fill=gray!7!white, inner sep=2.3pt, anchor=south east] at (axis cs: 5.5 -0.08, 3.5+0.15) {\tiny{$\nu_g = 1$}};
  	       
  	            \addplot[mark=diamond*, mark size=1.7pt, mark options={wind}, black] 
  	            table [x={x}, y={y}] {./plot/plot1/plot1a/area006_tie5_global2.txt}; 
	            \addplot[mark=otimes*, mark size=1.5pt, mark options={gas}, black] 
  	            table [x={x}, y={y}] {./plot/plot1/plot1a/area006_tie25_global2.txt}; 
  	            \addplot[mark=square*, mark size=1.5pt, mark options={biofuels}, black]
  	            table [x={x}, y={y}] {./plot/plot1/plot1a/area006_tie50_global2.txt}; 
  	            \draw [thin] (axis cs:10.5,0.1) -- (axis cs:10.5,180);
  	            \node[fill=gray!7!white, inner sep=2.3pt, anchor=south east] at (axis cs: 10.5 -0.08, 3.5+0.15) {\tiny{$\nu_g = 2$}};
  	            
  	            \addplot[mark=diamond*, mark size=1.7pt, mark options={wind}, black] 
  	            table [x={x}, y={y}] {./plot/plot1/plot1a/area006_tie5_global3.txt}; 
	            \addplot[mark=otimes*, mark size=1.5pt, mark options={gas}, black] 
  	            table [x={x}, y={y}] {./plot/plot1/plot1a/area006_tie25_global3.txt}; 
  	            \addplot[mark=square*, mark size=1.5pt, mark options={biofuels}, black]
  	            table [x={x}, y={y}] {./plot/plot1/plot1a/area006_tie50_global3.txt}; 
  	            \node[fill=gray!7!white, inner sep=2.3pt, anchor=south east] at (axis cs: 15.5 -0.08, 3.5+0.15) {\tiny{$\nu_g = 3$}};
	        \end{axis} 
	    \end{tikzpicture}}
	\end{tabular}\vspace{0.1cm}
	\captionsetup[subfigure]{oneside,margin={0.9cm,0cm}}
	\begin{tabular}{@{\hspace{-0.3cm}}c@{\hspace{0cm}}} \subfloat[]{\label{plot1b} \centering	
        \begin{tikzpicture}
	        \begin{axis} [box plot width=0.8mm, width=2.7cm, height=4.0cm,
	            grid style={line width=.3pt, draw=gray},
	            y tick label style={/pgf/number format/.cd, set thousands separator={},fixed},
	            xlabel style={align=center}, ylabel style ={yshift=-0.1cm},
    	        xlabel={Expected Number \\ of Tie Edges  $\bar{\gamma}_{c_i}$},
        	    ylabel={Iterations $\nu$},
  	            grid=major,   		
  	            xmin=0.5, xmax=2.5,	
  	            xtick={1, 2}, ytick={1300, 1483, 1666, 1850},
                xticklabels={5,50}, 
  	            tick label style={font=\scriptsize}, label style={font=\scriptsize},
  	            legend style={draw=black,fill=white,legend cell align=left,font=\tiny, at={(0.01,0.82)},anchor=west}]
	
	            \boxplot [forget plot, fill=wind, box plot whisker bottom index=1,
                box plot whisker top index=5, box plot box bottom index=2,
                box plot box top index=4, box plot median index=3] 
                {./plot/plot1/plot1b/syn_area006_tie5.txt};  
                
                \boxplot [forget plot, fill=biofuels, box plot whisker bottom index=1,
                box plot whisker top index=5, box plot box bottom index=2,
                box plot box top index=4, box plot median index=3] 
                {./plot/plot1/plot1b/syn_area006_tie50.txt}; 
	        \end{axis}
	    \end{tikzpicture}}
	\end{tabular}
	\captionsetup[subfigure]{oneside,margin={1.1cm,0cm}}
	\begin{tabular}{@{\hspace{-0.5cm}}c@{\hspace{0cm}}} \subfloat[]{\label{plot1c} \centering
        \begin{tikzpicture}
	        \begin{axis} [box plot width=0.8mm, ymode=log, width=6.6cm, height=4.0cm,
	            grid style={line width=.3pt, draw=gray},
	            xlabel style={align=center},
    	        xlabel={Local Iterations $\nu_l$ \\ Global Iterations $\nu_g = 1$},
        	    ylabel={Sequences $\nu_s$},
        	    ylabel style ={yshift=-0.1cm},
  	            grid=major,   		
  	            xmin=0, xmax=12, ymax = 600,
  	            ytick={10,100,1000},
  	            xtick={1,2,3,4,5, 7,8,9,10,11},
                xticklabels={1,5,30,60,90, 1,5,30,60,90}, 
  	            tick label style={font=\scriptsize}, label style={font=\scriptsize},
  	            legend style={draw=none, fill=gray!7!white, legend cell align=left, font=\tiny,
  	            at={(0.995, 0.985)}, anchor=north east, 
  	            {/tikz/every even column/.append style={column sep=0.1cm}}, 
  	            {/tikz/column 1/.append style={column sep=-0.15cm}},
  	            /tikz/column 4/.style={yshift=-0.025cm}, 
  	            /tikz/column 6/.style={yshift=-0.025cm}, 
  	            inner sep=0.5pt}, legend columns=4]
                
                \addlegendimage{legend image with text=$\bar{\gamma}_{c_i}$}
                \addlegendentry{}; 
	            \addlegendimage{mark=square*, only marks, mark options={wind}}
	            \addlegendentry{$5$}; 
	            \addlegendimage{mark=square*, only marks, mark options={biofuels}}
	            \addlegendentry{$50$}; 
	
	            \boxplot [forget plot, fill=wind, box plot whisker bottom index=1,
                box plot whisker top index=5, box plot box bottom index=2,
                box plot box top index=4, box plot median index=3] 
                {./plot/plot1/plot1c/dis_area006_tie5.txt};
                \draw [thin] (axis cs:6,0.1) -- (axis cs:6,1500);
                
                \boxplot [forget plot, fill=biofuels, box plot whisker bottom index=1,
                box plot whisker top index=5, box plot box bottom index=2,
                box plot box top index=4, box plot median index=3] 
                {./plot/plot1/plot1c/dis_area006_tie50.txt};
	        \end{axis}
	    \end{tikzpicture}}
	\end{tabular}
	\caption{The median values of the scale factor $\phi$ depending on the different local and global iteration schemes (subfigure a), the number of iterations of the synchronous GBP algorithm (subfigure b), and the number of sequences of the AGBP algorithm (subfigure c) for distributed system with $s = 2$, $\bar{\lambda}_{c_i} = 600$, $\bar{\gamma}_{c_i} = \{5, 25, 50\}$, $\delta = 0.01$.}
	\label{plot1}
\end{figure}

We start with a strictly diagonally dominant matrix $\mathbf{H}$ (i.e., the matrix is positive definite) obtained by using the diagonal increment $\delta > 0$, which guarantees the synchronous GBP convergence \cite{bickson}. We consider a distributed model with $s = 2$ clusters, where the internal matrices of dimension $m_{c_i} = 100$ create subgraphs with the expected number of internal edges equal to $\bar{\lambda}_{c_i} = 600$, $c_i \in \mathcal{C}$. We compare the performance of the synchronous GBP and AGBP algorithm for a different number of expected tie edges $\bar{\gamma}_{c_i} = \{5, 25, 50\}$. Finally, we set the diagonal increment to $\delta = 0.01$ to obtain the strictly diagonally dominant matrix $\mathbf{H}$.  
\figurename~\ref{plot1a} shows the median values of the scale factor $\phi$ depending on the different local and global iteration scheme strategies. Regardless of the number of global iterations $\nu_g$, as the number of tie edges ${\gamma}_{c_i}$ decreases, a larger number of local iterations $\nu_l$ is preferred. This expected outcome is a consequence of global iterations $\nu_g$ that have a minor impact on the estimate $\hat{\mathbf{x}}$ due to the small number of tie factor nodes. In addition, a decrease in the number of tie edges ${\gamma}_{c_i}$ leads to an increase in the value of the scale factor $\phi$, which is due to the decrease in the number of sequences $\nu_s$, as shown in \figurename~\ref{plot1c}. Comparing \figurename~\ref{plot1b} and \figurename~\ref{plot1c}, we note that the AGBP algorithm affects the convergence rate, leading to convergence in a significantly smaller number of iterations compared to synchronous GBP.  

\begin{figure}[ht]
	\centering \captionsetup[subfigure]{oneside,margin={1.0cm,0cm}}
	\begin{tabular}{@{\hspace{-0.2cm}}c@{}} 
	\subfloat[]{\label{plot2a} \centering	
        \begin{tikzpicture}
       \begin{axis}[width=9cm, height=4.0cm, ymode=log,
	            grid style={line width=.3pt, draw=gray},
                x tick label style={/pgf/number format/.cd, set thousands separator={},fixed},
                ylabel style ={yshift=-0.1cm},
  	            xlabel={Local Iterations $\nu_l$},
  	            ylabel={Scale Factor $\phi$},
  	            label style={font=\scriptsize},
  	            grid=major,
  	            ymin = 15, ymax = 2200, xmin = 0.5, xmax = 15.5,
  	            xtick={1,2,3,4,5, 6,7,8,9,10, 11,12,13,14,15},
  	            xticklabels={1,5,30,60,90, 2,5,30,60,90, 3,5,30,60,90},
  	            tick label style={font=\scriptsize},
  	            legend style={draw=none, fill=gray!7!white, legend cell align=left, font=\tiny,
  	            at={(0.995, 0.985)}, anchor=north east, 
  	            {/tikz/every even column/.append style={column sep=0.1cm}}, 
  	            {/tikz/column 1/.append style={column sep=-0.2cm}},
  	            /tikz/column 4/.style={yshift=-0.025cm}, 
  	            /tikz/column 6/.style={yshift=-0.025cm}, 
  	            /tikz/column 8/.style={yshift=-0.025cm}, inner sep=0.5pt}, 
  	            legend columns=4]
                
                \addlegendimage{legend image with text=$\bar{\gamma}_{c_i}$}
                \addlegendentry{}; 
	            \addlegendimage{mark=diamond*, only marks, mark options={wind}}
	            \addlegendentry{$5$}; 
                \addlegendimage{mark=otimes*, only marks, mark options={gas}}
	            \addlegendentry{$25$}; 
	            \addlegendimage{mark=square*, only marks, mark options={biofuels}}
	            \addlegendentry{$50$}; 

                \addplot[mark=diamond*, mark size=1.7pt, mark options={wind}, black] 
  	            table [x={x}, y={y}] {./plot/plot2/plot2a/area026_tie5_global1.txt}; 
	            \addplot[mark=otimes*, mark size=1.5pt, mark options={gas}, black] 
  	            table [x={x}, y={y}] {./plot/plot2/plot2a/area026_tie25_global1.txt}; 
  	            \addplot[mark=square*, mark size=1.5pt, mark options={biofuels}, black]
  	            table [x={x}, y={y}] {./plot/plot2/plot2a/area026_tie50_global1.txt}; 
  	            \draw [thin] (axis cs:5.5,0.1) -- (axis cs:5.5,2200);
  	            \node[fill=gray!7!white, inner sep=2.3pt, anchor=south east] at (axis cs: 5.5 -0.08, 15+1) {\tiny{$\nu_g = 1$}};
  	            
  	            \addplot[mark=diamond*, mark size=1.7pt, mark options={wind}, black] 
  	            table [x={x}, y={y}] {./plot/plot2/plot2a/area026_tie5_global2.txt}; 
	            \addplot[mark=otimes*, mark size=1.5pt, mark options={gas}, black] 
  	            table [x={x}, y={y}] {./plot/plot2/plot2a/area026_tie25_global2.txt}; 
  	            \addplot[mark=square*, mark size=1.5pt, mark options={biofuels}, black]
  	            table [x={x}, y={y}] {./plot/plot2/plot2a/area026_tie50_global2.txt}; 
  	            \draw [thin] (axis cs:10.5,0.1) -- (axis cs:10.5,2200);
  	            \node[fill=gray!7!white, inner sep=2.3pt, anchor=south east] at (axis cs: 10.5 -0.08, 15+1) {\tiny{$\nu_g = 2$}};
  	            
  	            \addplot[mark=diamond*, mark size=1.7pt, mark options={wind}, black] 
  	            table [x={x}, y={y}] {./plot/plot2/plot2a/area026_tie5_global3.txt}; 
	            \addplot[mark=otimes*, mark size=1.5pt, mark options={gas}, black] 
  	            table [x={x}, y={y}] {./plot/plot2/plot2a/area026_tie25_global3.txt}; 
  	            \addplot[mark=square*, mark size=1.5pt, mark options={biofuels}, black]
  	            table [x={x}, y={y}] {./plot/plot2/plot2a/area026_tie50_global3.txt}; 
  	            \node[fill=gray!7!white, inner sep=2.3pt, anchor=south east] at (axis cs: 15.5 -0.08, 15+1) {\tiny{$\nu_g = 3$}};
	        \end{axis} 
	    \end{tikzpicture}}
	\end{tabular}\vspace{0.1cm}
	\captionsetup[subfigure]{oneside,margin={0.9cm,0cm}}
	\begin{tabular}{@{\hspace{-0.3cm}}c@{\hspace{0cm}}} \subfloat[]{\label{plot2b} \centering	
        \begin{tikzpicture}
	        \begin{axis} [box plot width=0.8mm, width=2.7cm, height=4.0cm,
	            grid style={line width=.3pt, draw=gray},
	            y tick label style={/pgf/number format/.cd, set thousands separator={},fixed},
	            xlabel style={align=center}, ylabel style ={yshift=-0.1cm},
    	        xlabel={Expected Number \\ of Tie Edges  $\bar{\gamma}_{c_i}$},
        	    ylabel={Iterations $\nu$},
  	            grid=major,   		
  	            xmin=0.5, xmax=2.5,	
  	            xtick={1, 2}, ytick={5100, 5333, 5566, 5800},
                xticklabels={5,50}, 
  	            tick label style={font=\scriptsize}, label style={font=\scriptsize},
  	            legend style={draw=black,fill=white,legend cell align=left,font=\tiny, at={(0.01,0.82)},anchor=west}]
	
	            \boxplot [forget plot, fill=wind, box plot whisker bottom index=1,
                box plot whisker top index=5, box plot box bottom index=2,
                box plot box top index=4, box plot median index=3] 
                {./plot/plot2/plot2b/syn_area026_tie5.txt};  
                
                \boxplot [forget plot, fill=biofuels, box plot whisker bottom index=1,
                box plot whisker top index=5, box plot box bottom index=2,
                box plot box top index=4, box plot median index=3] 
                {./plot/plot2/plot2b/syn_area026_tie50.txt}; 
	        \end{axis}
	    \end{tikzpicture}}
	\end{tabular}
	\captionsetup[subfigure]{oneside,margin={1.1cm,0cm}}
	\begin{tabular}{@{\hspace{-0.5cm}}c@{\hspace{0cm}}} \subfloat[]{\label{plot2c} \centering
        \begin{tikzpicture}
	        \begin{axis} [box plot width=0.8mm, ymode=log, width=6.6cm, height=4.0cm,
	            grid style={line width=.3pt, draw=gray},
	            xlabel style={align=center},
    	        xlabel={Local Iterations $\nu_l$ \\ Global Iterations $\nu_g = 1$},
        	    ylabel={Sequences $\nu_s$},
        	    ylabel style ={yshift=-0.1cm},
  	            grid=major,   		
  	            xmin=0, xmax=12, ymax = 600,
  	            ytick={10,100,1000},
  	            xtick={1,2,3,4,5, 7,8,9,10,11},
                xticklabels={1,5,30,60,90, 1,5,30,60,90}, 
  	            tick label style={font=\scriptsize}, label style={font=\scriptsize},
  	            legend style={draw=none, fill=gray!7!white, legend cell align=left, font=\tiny,
  	            at={(0.995, 0.985)}, anchor=north east, 
  	            {/tikz/every even column/.append style={column sep=0.1cm}}, 
  	            {/tikz/column 1/.append style={column sep=-0.15cm}},
  	            /tikz/column 4/.style={yshift=-0.025cm}, 
  	            /tikz/column 6/.style={yshift=-0.025cm}, 
  	            inner sep=0.5pt}, legend columns=4]
                
                \addlegendimage{legend image with text=$\bar{\gamma}_{c_i}$}
                \addlegendentry{}; 
	            \addlegendimage{mark=square*, only marks, mark options={wind}}
	            \addlegendentry{$5$}; 
	            \addlegendimage{mark=square*, only marks, mark options={biofuels}}
	            \addlegendentry{$50$}; 
	
	            \boxplot [forget plot, fill=wind, box plot whisker bottom index=1,
                box plot whisker top index=5, box plot box bottom index=2,
                box plot box top index=4, box plot median index=3] 
                {./plot/plot2/plot2c/dis_area026_tie5.txt};
                \draw [thin] (axis cs:6,0.1) -- (axis cs:6,1500);
                
                \boxplot [forget plot, fill=biofuels, box plot whisker bottom index=1,
                box plot whisker top index=5, box plot box bottom index=2,
                box plot box top index=4, box plot median index=3] 
                {./plot/plot2/plot2c/dis_area026_tie50.txt};
	        \end{axis}
	    \end{tikzpicture}}
	\end{tabular}
	\caption{The median values of the scale factor $\phi$ depending on the different local and global iteration schemes (subfigure a), the number of iterations of the synchronous GBP algorithm (subfigure b), and the number of sequences of the AGBP algorithm (subfigure c) for distributed system with $s = 2$, $\bar{\lambda}_{c_i} = 2600$, $\bar{\gamma}_{c_i} = \{5, 25, 50\}$, $\delta = 0.01$.}
	\label{plot2}
\end{figure}
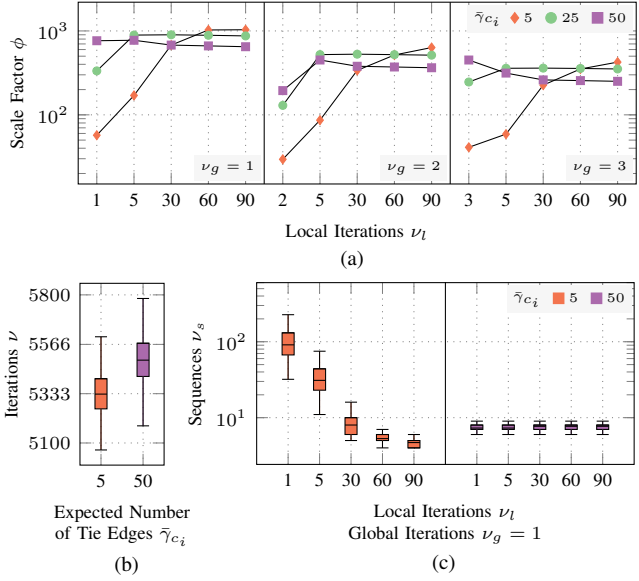

Next, we increase the expected number of internal edges to $\bar{\lambda}_{c_i} = 2600$, $c_i \in \mathcal{C}$. Similar to~\figurename~\ref{plot1a}, \figurename~\ref{plot2a} shows that the iteration scheme with $\nu_g = 1$ produces the highest values of the scale factor $\phi$ compared to the schemes with $\nu_g = 2$ and $\nu_g = 3$, regardless of the number of tie edges ${\gamma}_{c_i}$. Furthermore, comparing the same schemes for the increased number of internal edges ${\lambda}_{c_i}$ leads to even higher differences between the scale factors $\phi$. This is a consequence of local iterations $\nu_l$ that are becoming increasingly important in calculating the estimate $\hat{\mathbf{x}}$ as the value of ${\lambda}_{c_i}$ increases. In addition, the convergence rate of the synchronous GBP and AGBP algorithms has a significant impact on the scale factor $\phi$. According to \eqref{eqn:metric_time} and under condition $\lambda_{c_i} > \gamma_{c_i}$, it is intuitive to expect that the number of global iterations $\nu_g$ should be as small as possible (i.e., equal to one) to obtain the scaling factor $\phi$ as large as possible. 

As a conclusion, high values of the scale factor $\phi$ are caused by the fact that condition $\nu_s(\nu_g + \nu_l) < \nu$ holds. This inequality implies that the AGBP algorithm seeds a more rapid convergence rate, which can be explained by the fact that AGBP, compared to the synchronous GBP, decreases the spectral radius of the matrix affecting the evolution of the mean values. Comparing \figurename~\ref{plot1a} and \figurename~\ref{plot2a}, we observe an increase in the value of the scale factor $\phi$, as we increase the value of the expected number of internal edges $\lambda_{c_i}$. This increase improves the convergence rate of the AGBP algorithm, as shown in \figurename~\ref{plot1c} and \figurename~\ref{plot2c}. A similar trend for the convergence rate of the synchronous GBP algorithm is not observed in~\figurename~\ref{plot1b} and \figurename~\ref{plot2b}.

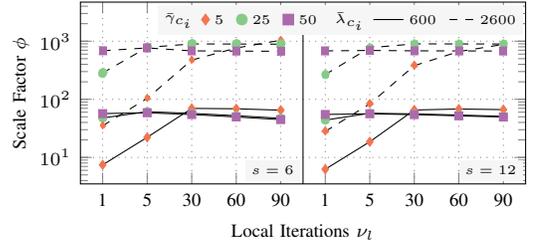
\begin{figure}[ht]
	\centering
	\begin{tikzpicture}
  	    \begin{axis}[width=7.5cm, height=4.0cm, ymode=log,
	        grid style={line width=.3pt, draw=gray},
            x tick label style={/pgf/number format/.cd, set thousands separator={},fixed},
            ylabel style ={yshift=-0.1cm},
  	        xlabel={Local Iterations $\nu_l$},
  	        ylabel={Scale Factor $\phi$},
  	        label style={font=\scriptsize},
  	        grid=major,
  	        ymin = 3.5, ymax = 4700, xmin = 0.5, xmax = 10.5,
  	        xtick={1,2,3,4,5, 6,7,8,9,10},
  	        xticklabels={1,5,30,60,90, 1,5,30,60,90},
  	        tick label style={font=\scriptsize},
    	    legend style={draw=none, fill=gray!7!white, legend cell align=left, font=\tiny,
  	        at={(0.995, 0.985)}, anchor=north east, 
  	        {/tikz/every even column/.append style={column sep=0.1cm}}, 
  	        {/tikz/column 1/.append style={column sep=-0.2cm}},
  	        /tikz/column 4/.style={yshift=-0.025cm}, 
  	        /tikz/column 6/.style={yshift=-0.025cm}, 
  	        /tikz/column 8/.style={yshift=-0.025cm}, 
  	        {/tikz/column 9/.append style={column sep=-0.2cm}},
  	        /tikz/column 12/.style={yshift=-0.025cm}, 
  	        /tikz/column 14/.style={yshift=-0.025cm}, 
  	        {/tikz/column 13/.append style={column sep=-0.07cm}},
  	        inner sep=0.5pt}, legend columns=10]
                
            \addlegendimage{legend image with text=$\bar{\gamma}_{c_i}$}
            \addlegendentry{}; 
	        \addlegendimage{mark=diamond*, only marks, mark options={wind}}
	        \addlegendentry{$5$}; 
            \addlegendimage{mark=otimes*, only marks, mark options={gas}}
	        \addlegendentry{$25$}; 
	        \addlegendimage{mark=square*, only marks, mark options={biofuels}}
	        \addlegendentry{$50$}; 
	        \addlegendimage{legend image with text=$\bar{\lambda}_{c_i}$}
	        \addlegendentry{};
	        \addlegendimage{legend image code/.code={\draw[black] plot coordinates {(0cm,0cm) (0.4cm,0cm)};}}
	        \addlegendentry{$600$}; 
	        \addlegendimage{legend image code/.code={\draw[black, dashed] plot coordinates {(0cm,0cm) (0.4cm,0cm)};}}
	        \addlegendentry{$2600$}; 
            
            \addplot[mark=diamond*, mark size=1.7pt, mark options={wind}, black] 
  	        table [x={x}, y={y}] {./plot/plot3/area006_tie5_cluster6_global1.txt};
	        \addplot[mark=otimes*, mark size=1.5pt, mark options={gas}, black] 
  	        table [x={x}, y={y}] {./plot/plot3/area006_tie25_cluster6_global1.txt}; 
	        \addplot[mark=square*, mark size=1.5pt, mark options={biofuels}, black] 
  	        table [x={x}, y={y}] {./plot/plot3/area006_tie50_cluster6_global1.txt}; 
  	       
  	        \addplot[mark=diamond*, mark size=1.5pt, mark options={wind}, black, dashed] 
  	        table [x={x}, y={y}] {./plot/plot3/area026_tie5_cluster6_global1.txt}; 
  	        \addplot[mark=otimes*, mark size=1.7pt, mark options={gas}, black, dashed] 
  	        table [x={x}, y={y}] {./plot/plot3/area026_tie25_cluster6_global1.txt}; 
	        \addplot[mark=square*, mark size=1.5pt, mark options={biofuels}, black, dashed] 
  	        table [x={x}, y={y}] {./plot/plot3/area026_tie50_cluster6_global1.txt}; 
  	        
  	        \draw [thin] (axis cs:5.5,0.1) -- (axis cs:5.5,4000);
  	        \node[fill=gray!7!white, inner sep=2.3pt, anchor=south east] at (axis cs: 5.5-0.06, 3.5+0.4) {\tiny{$s = 6$}};

  	        \addplot[mark=diamond*, mark size=1.7pt, mark options={wind}, black] 
  	        table [x={x}, y={y}] {./plot/plot3/area006_tie5_cluster12_global1.txt}; 
	        \addplot[mark=otimes*, mark size=1.5pt, mark options={gas}, black] 
  	        table [x={x}, y={y}] {./plot/plot3/area006_tie25_cluster12_global1.txt}; 
  	        \addplot[mark=square*, mark size=1.5pt, mark options={biofuels}, black]
  	        table [x={x}, y={y}] {./plot/plot3/area006_tie50_cluster12_global1.txt};
  	        
  	        \addplot[mark=diamond*, mark size=1.7pt, mark options={wind}, black, dashed] 
  	        table [x={x}, y={y}] {./plot/plot3/area026_tie5_cluster12_global1.txt}; 
	        \addplot[mark=otimes*, mark size=1.5pt, mark options={gas}, black, dashed] 
  	        table [x={x}, y={y}] {./plot/plot3/area026_tie25_cluster12_global1.txt}; 
  	        \addplot[mark=square*, mark size=1.5pt, mark options={biofuels}, black, dashed]
  	        table [x={x}, y={y}] {./plot/plot3/area026_tie50_cluster12_global1.txt}; 
  	      
  	        \node[fill=gray!7!white, inner sep=2.3pt, anchor=south east] at (axis cs: 10.5-0.06, 3.5+0.4) {\tiny{$s = 12$}};	    
  	    \end{axis}
	\end{tikzpicture}
	\caption{The median values of the scale factor $\phi$ depending on the different local iterations $\nu_l$, where the number of global iterations is equal to $\nu_g = 1$, for distributed systems with $s = \{6,12\}$, $\bar{\lambda}_{c_i} = \{600, 2600\}$, $\bar{\gamma}_{c_i} = \{5,25,50\}$, $\delta = 0.01$.}
	\label{plot3}
\end{figure}

To investigate the scalability of the AGBP algorithm, we increase the number of clusters $s$, where we kept the expected number of tie edges per each cluster as before $\bar{\gamma}_{c_i} = \{5, 25, 50\}$. Compared with \figurename~\ref{plot1a} and \figurename~\ref{plot2a}, \figurename~\ref{plot3} shows that the scale factor $\phi$ remains at almost the same maximum value regardless of the number of clusters $s$. This effect is caused by the fact that both the number of iterations $\nu$ and the number of sequences $\nu_s$, remain almost unchanged compared to the scenario with $s = 2$, whereby $\nu$ is significantly greater than $\kappa\nu_s(\nu_g + \nu_l)$. Hence, according to \eqref{eqn:metric_time}, the time $\tau_c$ increases the overall value as the number of clusters $s$ increases (this follows from increase of the processing time $\tau_m$).

To further test scalability, we analyse distributed systems with $s=12$, where the systems are formed according to $\bar{\lambda}_{c_i} = \{1800, 7800\}$, $\bar{\gamma}_{c_i} = \{15,75,150\}$, and $m_{c_i} = 300$, $c_i \in \mathcal{C}$. Here, we increase the number of internal and tie edges, as well as the number of variable and factor nodes by a factor of three compared to the previous scenario. Increasing the number of nodes and edges does not affect the maximum value of the scale factor $\phi$, while again the number of iterations $\nu$, and the number of sequences $\nu_s$, did not increase the values compared to the cases analysed previously. Finally, we conclude that both the number of iterations $\nu$, and the number of sequences $\nu_s$, depends primarily on the number of internal edges ${\lambda}_{c_i}$, while being independent of the size of the factor graph or the number of clusters $s$.

\begin{figure}[ht]
	\centering
	\begin{tikzpicture}
  	    \begin{axis}[width=7.5cm, height=4.0cm,
	        grid style={line width=.3pt, draw=gray},
            x tick label style={/pgf/number format/.cd, set thousands separator={},fixed},
            ylabel style ={yshift=-0.1cm},
  	        xlabel={Local Iterations $\nu_l$},
  	        ylabel={Scale Factor $\phi$},
  	        label style={font=\scriptsize},
  	        grid=major,
  	        ymin = -110, ymax = 160, 
  	        xmin = 0.5, xmax = 10.5,
  	        xtick={1,2,3,4,5, 6,7,8,9,10},
  	        ytick={-60,0,60,120},
  	        xticklabels={1,5,30,60,90, 1,5,30,60,90},
  	        tick label style={font=\scriptsize},
    	    legend style={draw=none, fill=gray!7!white, legend cell align=left, font=\tiny,
  	        at={(0.995, 0.985)}, anchor=north east, 
  	        {/tikz/every even column/.append style={column sep=0.1cm}}, 
  	        {/tikz/column 1/.append style={column sep=-0.2cm}},
  	        /tikz/column 4/.style={yshift=-0.025cm}, 
  	        /tikz/column 6/.style={yshift=-0.025cm}, 
  	        {/tikz/column 7/.append style={column sep=-0.2cm}},
  	        /tikz/column 10/.style={yshift=-0.025cm}, 
  	        /tikz/column 12/.style={yshift=-0.025cm}, 
  	        {/tikz/column 11/.append style={column sep=-0.07cm}},
  	        inner sep=0.5pt}, legend columns=10]
                
            \addlegendimage{legend image with text=$\bar{\gamma}_{c_i}$}
            \addlegendentry{}; 
	        \addlegendimage{mark=diamond*, only marks, mark options={wind}}
	        \addlegendentry{$5$}; 
	        \addlegendimage{mark=square*, only marks, mark options={biofuels}}
	        \addlegendentry{$50$}; 
	        \addlegendimage{legend image with text=$\bar{\lambda}_{c_i}$}
	        \addlegendentry{};
	        \addlegendimage{legend image code/.code={\draw[black] plot coordinates {(0cm,0cm) (0.4cm,0cm)};}}
	        \addlegendentry{$600$}; 
	        \addlegendimage{legend image code/.code={\draw[black, dashed] plot coordinates {(0cm,0cm) (0.4cm,0cm)};}}
	        \addlegendentry{$2600$}; 
            
            \addplot[mark=diamond*, mark size=1.7pt, mark options={wind}, black] 
  	        table [x={x}, y={y}] {./plot/plot4/area006_tie5_delta01.txt};
	        \addplot[mark=square*, mark size=1.5pt, mark options={biofuels}, black] 
  	        table [x={x}, y={y}] {./plot/plot4/area006_tie50_delta01.txt};
  	       
  	        \addplot[mark=diamond*, mark size=1.5pt, mark options={wind}, black, dashed] 
  	        table [x={x}, y={y}] {./plot/plot4/area026_tie5_delta01.txt};
	        \addplot[mark=square*, mark size=1.5pt, mark options={biofuels}, black, dashed] 
  	        table [x={x}, y={y}] {./plot/plot4/area026_tie50_delta01.txt};
  	        
  	        \draw [thin] (axis cs:5.5,-110) -- (axis cs:5.5,150);
  	        \node[fill=gray!7!white, inner sep=2.3pt, anchor=south east] at (axis cs: 5.5-0.06, -110+4) {\tiny{$\delta = 0.1$}};
  	        
  	        \addplot[mark=diamond*, mark size=1.7pt, mark options={wind}, black] 
  	        table [x={x}, y={y}] {./plot/plot4/area006_tie5_delta1.txt};
  	        \addplot[mark=square*, mark size=1.5pt, mark options={biofuels}, black]
  	        table [x={x}, y={y}] {./plot/plot4/area006_tie50_delta1.txt};
  	        
  	        \addplot[mark=diamond*, mark size=1.7pt, mark options={wind}, black, dashed] 
  	        table [x={x}, y={y}] {./plot/plot4/area026_tie5_delta1.txt};
  	        \addplot[mark=square*, mark size=1.5pt, mark options={biofuels}, black, dashed]
  	        table [x={x}, y={y}] {./plot/plot4/area026_tie50_delta1.txt};
  	      
 	        \node[fill=gray!7!white, inner sep=2.3pt, anchor=south east] at (axis cs: 10.5-0.06, -110+4) {\tiny{$\delta = 1$}};
	    \end{axis}
	\end{tikzpicture}
	\caption{The median values of the scale factor $\phi$ depending on the different local iterations $\nu_l$, where the number of global iterations is equal to $\nu_g = 1$, for distributed systems with $s = 2$, $\bar{\lambda}_{c_i} = \{600, 2600\}$, $\bar{\gamma}_{c_i} = \{5,50\}$, $\delta = \{0.1,1\}$.}
	\label{plot4}
\end{figure}
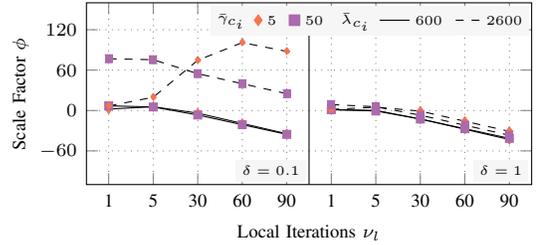

Finally, we consider the influence of the diagonal increment $\delta$ on the scale factor $\phi$. Comparing \figurename~\ref{plot4} with \figurename~\ref{plot1a} and \figurename~\ref{plot2a} for $\nu_g = 1$, we observe a decrease in the value of the scale factor $\phi$ with increasing the diagonal increment $\delta$. As expected, increasing the diagonal loading causes a more rapid convergence rate of both algorithms, having more significant effect on the synchronous GBP algorithm. Hence, for certain sequences of local and global iterations, condition $\nu < \nu_s(\nu_g + \nu_l)$ is satisfied, resulting in $\phi < 0$. More precisely, for $\phi < 0$ the AGBP will always produce the estimate $\hat{\mathbf{x}}$ for a time period exceeding that of the synchronous GBP algorithm.

Consider now a symmetric diagonally dominant matrix $\mathbf{H}$ (i.e., the matrix is positive semidefinite) obtained using the diagonal increment equal to $\delta =0$. For the baseline scenario, $s = 2$, $\bar{\lambda}_{c_i} = \{600, 2600\}$, $\bar{\gamma}_{c_i} = \{5, 25, 50\}$, $m_{c_i} = 100$, the synchronous GBP algorithm for $\bar{\lambda}_{c_i} = 600$ converges with probability $\{0.39, 0.34, 0.28\}$ depending on the expected number of tie edges $\{5, 25, 50\}$, respectively, while for $\bar{\lambda}_{c_i} = 2600$ does not converge at all. In contrast, the AGBP algorithm always converges for any set of distributed model’s parameters. Also, switching from the strictly diagonally dominant matrix to the diagonally dominant matrix did not cause significant changes in the number of sequences $\nu_s$.

\subsection{Linear Models with Nonsymmetric Matrices Properties} \label{nonsymmetric}
\begin{figure}[ht]
	\centering 	\captionsetup[subfigure]{oneside,margin={0.9cm,0cm}}
	\begin{tabular}{@{\hspace{-0.3cm}}c@{\hspace{0cm}}} \subfloat[]{\label{plot6a} \centering	
        \begin{tikzpicture}
	        \begin{axis} [box plot width=0.8mm, width=2.7cm, height=4.0cm,
	            grid style={line width=.3pt, draw=gray},
	            y tick label style={/pgf/number format/.cd, set thousands separator={},fixed},
	            xlabel style={align=center}, ylabel style ={yshift=-0.1cm},
    	        xlabel={Expected Number \\ of Tie Edges  $\bar{\gamma}_{c_i}$},
        	    ylabel={Iterations $\nu$},
  	            grid=major,   		
  	            xmin=0.5, xmax=2.5,	
  	            xtick={1, 2},
  	            ytick={400,1100,1800,2500},
                xticklabels={5,50}, 
  	            tick label style={font=\scriptsize}, label style={font=\scriptsize},
  	            legend style={draw=black,fill=white,legend cell align=left,font=\tiny, at={(0.01,0.82)},anchor=west}]
	
	            \boxplot [forget plot, fill=wind, box plot whisker bottom index=1,
                box plot whisker top index=5, box plot box bottom index=2,
                box plot box top index=4, box plot median index=3] 
                {./plot/plot6/plot6a/syn_area006_tie5.txt};  
                
                \boxplot [forget plot, fill=biofuels, box plot whisker bottom index=1,
                box plot whisker top index=5, box plot box bottom index=2,
                box plot box top index=4, box plot median index=3] 
                {./plot/plot6/plot6a/syn_area006_tie50.txt}; 
	        \end{axis}
	    \end{tikzpicture}}
	\end{tabular}
	\captionsetup[subfigure]{oneside,margin={1.1cm,0cm}}
	\begin{tabular}{@{\hspace{-0.5cm}}c@{\hspace{0cm}}} \subfloat[]{\label{plot6b} \centering
        \begin{tikzpicture}
	        \begin{axis} [box plot width=0.8mm, ymode=log, width=6.6cm, height=4.0cm,
	            grid style={line width=.3pt, draw=gray},
	            xlabel style={align=center},
    	        xlabel={Local Iterations $\nu_l$ \\ Global Iterations $\nu_g = 1$},
        	    ylabel={Sequences $\nu_s$},
        	    ylabel style ={yshift=-0.1cm},
  	            grid=major,   		
  	            xmin=0, xmax=12, ymax = 600,
  	            ytick={10,100,1000},
  	            xtick={1,2,3,4,5, 7,8,9,10,11},
                xticklabels={1,5,30,60,90, 1,5,30,60,90}, 
  	            tick label style={font=\scriptsize}, label style={font=\scriptsize},
  	            legend style={draw=none, fill=gray!7!white, legend cell align=left, font=\tiny,
  	            at={(0.995, 0.985)}, anchor=north east, 
  	            {/tikz/every even column/.append style={column sep=0.1cm}}, 
  	            {/tikz/column 1/.append style={column sep=-0.15cm}},
  	            /tikz/column 4/.style={yshift=-0.025cm}, 
  	            /tikz/column 6/.style={yshift=-0.025cm}, 
  	            inner sep=0.5pt}, legend columns=4]
                
                \addlegendimage{legend image with text=$\bar{\gamma}_{c_i}$}
                \addlegendentry{}; 
	            \addlegendimage{mark=square*, only marks, mark options={wind}}
	            \addlegendentry{$5$}; 
	            \addlegendimage{mark=square*, only marks, mark options={biofuels}}
	            \addlegendentry{$50$}; 
	
	            \boxplot [forget plot, fill=wind, box plot whisker bottom index=1,
                box plot whisker top index=5, box plot box bottom index=2,
                box plot box top index=4, box plot median index=3] 
                {./plot/plot6/plot6b/dis_area006_tie5.txt};
                \draw [thin] (axis cs:6,0.1) -- (axis cs:6,1500);
                
                \boxplot [forget plot, fill=biofuels, box plot whisker bottom index=1,
                box plot whisker top index=5, box plot box bottom index=2,
                box plot box top index=4, box plot median index=3] 
                {./plot/plot6/plot6b/dis_area006_tie50.txt};
	        \end{axis}
	    \end{tikzpicture}}
	\end{tabular}\vspace{0.1cm}
	\captionsetup[subfigure]{oneside,margin={0.9cm,0cm}}
	\begin{tabular}{@{\hspace{-0.3cm}}c@{\hspace{0cm}}} \subfloat[]{\label{plot6c} \centering	
        \begin{tikzpicture}
	        \begin{axis} [box plot width=0.8mm, width=2.7cm, height=4.0cm,
	            grid style={line width=.3pt, draw=gray},
	            y tick label style={/pgf/number format/.cd, set thousands separator={},fixed},
	            xlabel style={align=center}, ylabel style ={yshift=-0.1cm},
    	        xlabel={Expected Number \\ of Tie Edges $\bar{\gamma}_{c_i}$},
        	    ylabel={Iterations $\nu$},
  	            grid=major,   		
  	            xmin=0.5, xmax=2.5, ymax = 9999,	
  	            xtick={1, 2},
  	            ytick={9000,9300,9600,9900},
                xticklabels={5,50}, 
  	            tick label style={font=\scriptsize}, label style={font=\scriptsize},
  	            legend style={draw=black,fill=white,legend cell align=left,font=\tiny, at={(0.01,0.82)},anchor=west}]
	
	            \boxplot [forget plot, fill=wind, box plot whisker bottom index=1,
                box plot whisker top index=5, box plot box bottom index=2,
                box plot box top index=4, box plot median index=3] 
                {./plot/plot6/plot6c/syn_area026_tie5.txt};  
                
                \boxplot [forget plot, fill=biofuels, box plot whisker bottom index=1,
                box plot whisker top index=5, box plot box bottom index=2,
                box plot box top index=4, box plot median index=3] 
                {./plot/plot6/plot6c/syn_area026_tie50.txt}; 
	        \end{axis}
	    \end{tikzpicture}}
	\end{tabular}
	\captionsetup[subfigure]{oneside,margin={1.1cm,0cm}}
	\begin{tabular}{@{\hspace{-0.5cm}}c@{\hspace{0cm}}} \subfloat[]{\label{plot6d} \centering
        \begin{tikzpicture}
	        \begin{axis} [box plot width=0.8mm, ymode=log, width=6.6cm, height=4.0cm,
	            grid style={line width=.3pt, draw=gray},
	            xlabel style={align=center},
    	        xlabel={Local Iterations $\nu_l$ \\ Global Iterations $\nu_g = 1$},
        	    ylabel={Sequences $\nu_s$},
        	    ylabel style ={yshift=-0.1cm},
  	            grid=major,   		
  	            xmin=0, xmax=12, ymax = 600,
  	            ytick={10,100,1000},
  	            xtick={1,2,3,4,5, 7,8,9,10,11},
                xticklabels={1,5,30,60,90, 1,5,30,60,90}, 
  	            tick label style={font=\scriptsize}, label style={font=\scriptsize},
  	            legend style={draw=none, fill=gray!7!white, legend cell align=left, font=\tiny,
  	            at={(0.995, 0.985)}, anchor=north east, 
  	            {/tikz/every even column/.append style={column sep=0.1cm}}, 
  	            {/tikz/column 1/.append style={column sep=-0.15cm}},
  	            /tikz/column 4/.style={yshift=-0.025cm}, 
  	            /tikz/column 6/.style={yshift=-0.025cm}, 
  	            inner sep=0.5pt}, legend columns=4]
                
                \addlegendimage{legend image with text=$\bar{\gamma}_{c_i}$}
                \addlegendentry{}; 
	            \addlegendimage{mark=square*, only marks, mark options={wind}}
	            \addlegendentry{$5$}; 
	            \addlegendimage{mark=square*, only marks, mark options={biofuels}}
	            \addlegendentry{$50$}; 
	
	            \boxplot [forget plot, fill=wind, box plot whisker bottom index=1,
                box plot whisker top index=5, box plot box bottom index=2,
                box plot box top index=4, box plot median index=3] 
                {./plot/plot6/plot6d/dis_area026_tie5.txt};
                \draw [thin] (axis cs:6,0.1) -- (axis cs:6,1500);
                
                \boxplot [forget plot, fill=biofuels, box plot whisker bottom index=1,
                box plot whisker top index=5, box plot box bottom index=2,
                box plot box top index=4, box plot median index=3] 
                {./plot/plot6/plot6d/dis_area026_tie50.txt};
	        \end{axis}
	    \end{tikzpicture}}
	\end{tabular}
	\caption{The number of iterations of the synchronous GBP algorithm for $\Lambda_{c_i} = 0.06$ (subfigure a) and $\Lambda_{c_i} = 0.26$ (subfigure c), and the number of sequences of the AGBP algorithm for $\bar{\lambda}_{c_i} = 600$ (subfigure b) and $\bar{\lambda}_{c_i} = 2600$ (subfigure d) for distributed system with $s = 2$, $\bar{\gamma}_{c_i} = \{5, 50\}$, $\delta = 0.01$.}
	\label{plot6}
\end{figure}
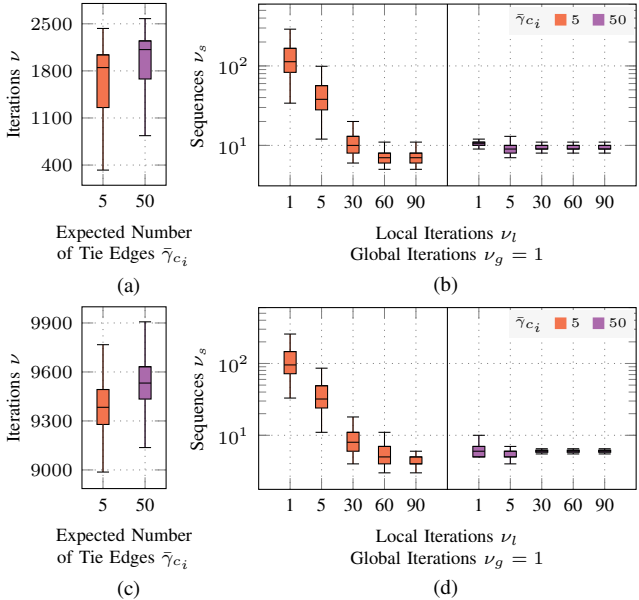

We complement results in Section \ref{symmetric} by analysing distributed models where each cluster consists of the nonsymmetric matrix $\mathbf{H}_{c_i}$, and matrices $\mathbf{H}_{c_i,c_j}$, $m_{c_i} = n_{c_i} = n_{c_j}$. This setup forms the nonsymmetric matrix $\mathbf{H} \in \mathbb{R}^{m \times n}$, $m=n$. In comparison with \figurename~{\ref{plot1b}, \figurename~\ref{plot6a} shows a significant increase in the number of synchronous GBP iterations $\nu$. However, the trend for the AGBP algorithm is the opposite, with the number of sequences $\nu_s$ decreasing as illustrated in~\figurename~\ref{plot6b} and \figurename~\ref{plot1c}. Similar observations hold for the scenarios depicted in \figurename~\ref{plot2b}, \figurename~\ref{plot2c}}, compared to \figurename~\ref{plot6c}, and \figurename~\ref{plot6d}.

Thus, the nonsymmetric matrix properties slow down convergence rates of the synchronous GBP algorithm, while causing more rapid convergence rates of the AGBP algorithm compared with the systems with symmetric matrix properties. 
\vspace{-0.36cm}

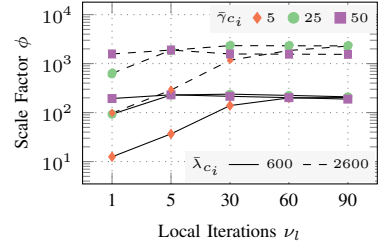
\begin{figure}[ht]
	\centering	
        \begin{tikzpicture}
    	    \begin{axis}[width=5.5cm, height=4.0cm, ymode=log,
	        grid style={line width=.3pt, draw=gray},
            x tick label style={/pgf/number format/.cd, set thousands separator={},fixed},
            ylabel style ={yshift=-0.1cm},
  	        xlabel={Local Iterations $\nu_l$},
  	        ylabel={Scale Factor $\phi$},
  	        label style={font=\scriptsize},
  	        grid=major,
  	        ymin = 3.5, ymax = 18000, xmin = 0.5, xmax = 5.5,
  	        xtick={1,2,3,4,5},
  	        xticklabels={1,5,30,60,90},
  	        tick label style={font=\scriptsize},
    	    legend style={draw=none, fill=gray!7!white, legend cell align=left, font=\tiny,
  	        at={(0.995, 0.985)}, anchor=north east, 
  	        {/tikz/every even column/.append style={column sep=0.1cm}}, 
  	        {/tikz/column 1/.append style={column sep=-0.2cm}},
  	        /tikz/column 4/.style={yshift=-0.025cm}, 
  	        /tikz/column 6/.style={yshift=-0.025cm}, 
  	        /tikz/column 8/.style={yshift=-0.025cm}, 
  	        {/tikz/column 9/.append style={column sep=-0.2cm}},
  	        inner sep=0.5pt}, legend columns=10]
                
            \addlegendimage{legend image with text=$\bar{\gamma}_{c_i}$}
            \addlegendentry{}; 
	        \addlegendimage{mark=diamond*, only marks, mark options={wind}}
	        \addlegendentry{$5$}; 
            \addlegendimage{mark=otimes*, only marks, mark options={gas}}
	        \addlegendentry{$25$}; 
	        \addlegendimage{mark=square*, only marks, mark options={biofuels}}
	        \addlegendentry{$50$}; 
            
            \addplot[mark=diamond*, mark size=1.7pt, mark options={wind}, black] 
  	        table [x={x}, y={y}] {./plot/plot7/area006_tie5_global1.txt};
	        \addplot[mark=otimes*, mark size=1.5pt, mark options={gas}, black] 
  	        table [x={x}, y={y}] {./plot/plot7/area006_tie25_global1.txt};
	        \addplot[mark=square*, mark size=1.5pt, mark options={biofuels}, black] 
  	        table [x={x}, y={y}] {./plot/plot7/area006_tie50_global1.txt};
  	       
  	        \addplot[mark=diamond*, mark size=1.5pt, mark options={wind}, black, dashed] 
  	        table [x={x}, y={y}] {./plot/plot7/area026_tie5_global1.txt};
  	        \addplot[mark=otimes*, mark size=1.7pt, mark options={gas}, black, dashed] 
  	        table [x={x}, y={y}] {./plot/plot7/area026_tie25_global1.txt};
	        \addplot[mark=square*, mark size=1.5pt, mark options={biofuels}, black, dashed] 
  	        table [x={x}, y={y}] {./plot/plot7/area026_tie50_global1.txt};

            \node[fill=gray!7!white, inner sep=2.3pt] at (axis cs: 3.855, 8) 
            {\tiny{$\bar{\lambda}_{c_i}$}\quad\quad\quad\;\,\tiny{$600$}\quad\quad\quad\tiny{$2600$}};
            
            \draw [black] (axis cs:2.98,8.7) -- (axis cs:3.49,8.7);
            \draw [black, dashed] (axis cs:4.24,8.7) -- (axis cs:4.75,8.7);
	    \end{axis}
	    \end{tikzpicture}
	\caption{The median values of the scale factor $\phi$ depending on the different local iterations $\nu_l$, where the number of global iterations is equal to $\nu_g = 1$, for distributed systems with $s = 2$, $\bar{\lambda}_{c_i} = \{600, 2600\}$, $\bar{\gamma}_{c_i} = \{5,25,50\}$, $\delta = 0.01$.}
	\label{plot7}
\end{figure}

With these observations, the scaling factor $\phi$ is higher for distributed systems with nonsymmetric matrices compared to systems with symmetric matrices, as illustrated in \figurename~\ref{plot7}. The trend for the scaling factor is similar across different configurations. We note that the statements given in Section \ref{symmetric} are also valid for distributed systems consisting of nonsymmetric matrices.

\subsection{Linear Models with Rectangular Matrices} \label{rectangular}
After the analysis of linear models with square matrices, we proceed by analysing linear models consisting of rectangular matrices. In particular, a rectangular matrix $\mathbf{H}$, $m > n$, consists of the matrices $\mathbf{H}_{c_i}$ and $\mathbf{H}_{c_i,c_j}$ that are also rectangular $m_{c_i} > n_{c_i}$, $n_{c_i} = n_{c_j}$. We start by defining exactly $n_{c_i}$ linearly independent equations, followed by adding a set of $m_{c_i} - n_{c_i}$ dependent equations to form the internal matrix $\mathbf{H}_{c_i}$. Thus, we reuse square nonsymmetric matrices from Section \ref{nonsymmetric} with $\delta = 0.01$, which are extended with $m_{c_i} - n_{c_i}$ additional rows. Each additional row of the internal matrix $\mathbf{H}_{c_i}$ or the tie matrix $\mathbf{H}_{c_i,c_j}$ defines as many expected edges as one row defines for the case of the internal or tie matrix in Section \ref{nonsymmetric}. 

\begin{figure}[ht] 
	\centering \captionsetup[subfigure]{oneside,margin={1.0cm,0cm}}
	\begin{tabular}{@{\hspace{-0.2cm}}c@{}} 
	\subfloat[]{\label{plot8a} \centering	
      \begin{tikzpicture}
      \begin{axis}[width=9cm, height=4.0cm,
	            grid style={line width=.3pt, draw=gray},
                x tick label style={/pgf/number format/.cd, set thousands separator={},fixed},
                ylabel style ={yshift=-0.1cm},
  	            xlabel={Local Iterations $\nu_l$},
  	            ylabel={Scale Factor $\phi$},
  	            label style={font=\scriptsize},
  	            grid=major,
  	            ymin = -40, ymax = 130, 
  	            xmin = 0.5, xmax = 15.5,
  	            xtick={1,2,3,4,5, 6,7,8,9,10, 11,12,13,14,15},
  	            xticklabels={1,5,30,60,90, 1,2,3,4,5, 1,2,3,4,5},
  	            tick label style={font=\scriptsize},
  	            legend style={draw=none, fill=gray!7!white, legend cell align=left, font=\tiny,
  	            at={(0.995, 0.985)}, anchor=north east, 
  	            {/tikz/every even column/.append style={column sep=0.1cm}}, 
  	            {/tikz/column 1/.append style={column sep=-0.2cm}},
  	            /tikz/column 4/.style={yshift=-0.025cm}, 
  	            /tikz/column 6/.style={yshift=-0.025cm}, 
  	            /tikz/column 8/.style={yshift=-0.025cm}, inner sep=0.5pt}, 
  	            legend columns=4]
                
                \addlegendimage{legend image with text=$\bar{\gamma}_{c_i}$}
                \addlegendentry{}; 
	            \addlegendimage{mark=diamond*, only marks, mark options={wind}}
	            \addlegendentry{$0.05m_{c_i}$}; 
                \addlegendimage{mark=otimes*, only marks, mark options={gas}}
	            \addlegendentry{$0.25m_{c_i}$}; 
	            \addlegendimage{mark=square*, only marks, mark options={biofuels}}
	            \addlegendentry{$0.5m_{c_i}$}; 
	            
                \addplot[mark=diamond*, mark size=1.7pt, mark options={wind}, black] 
  	            table [x={x}, y={y}] {./plot/plot8/plot8a/tie005_row120.txt}; 
  	            \addplot[mark=diamond*, mark size=1.7pt, mark options={wind}, black, dashed] 
  	            table [x={x}, y={y}] {./plot/plot8/plot8a/tie005_row200.txt}; 
  	            \addplot[mark=diamond*, mark size=1.7pt, mark options={wind}, black, densely dash dot dot] 
  	            table [x={x}, y={y}] {./plot/plot8/plot8a/tie005_row300.txt}; 
  	            \draw [thin] (axis cs:5.5,-40) -- (axis cs:5.5,140);  	            
  
	            \addplot[mark=otimes*, mark size=1.5pt, mark options={gas}, black] 
  	            table [x={x}, y={y}] {./plot/plot8/plot8a/tie025_row120.txt}; 
  	           	\addplot[mark=otimes*, mark size=1.5pt, mark options={gas}, black, dashed] 
  	            table [x={x}, y={y}] {./plot/plot8/plot8a/tie025_row200.txt};
  	            \addplot[mark=otimes*, mark size=1.7pt, mark options={gas}, black, densely dash dot dot] 
  	            table [x={x}, y={y}] {./plot/plot8/plot8a/tie025_row300.txt};
  	            \draw [thin] (axis cs:10.5,-20) -- (axis cs:10.5,120);
  	            
  	            \addplot[mark=square*, mark size=1.5pt, mark options={biofuels}, black]
  	            table [x={x}, y={y}] {./plot/plot8/plot8a/tie05_row120.txt}; 
  	           	\addplot[mark=square*, mark size=1.5pt, mark options={biofuels}, black, dashed] 
  	            table [x={x}, y={y}] {./plot/plot8/plot8a/tie05_row200.txt};  	            
  	            \addplot[mark=square*, mark size=1.7pt, mark options={biofuels}, black, densely dash dot dot] 
  	            table [x={x}, y={y}] {./plot/plot8/plot8a/tie05_row300.txt};
  	            
  	            \node[fill=gray!7!white, inner sep=2.3pt] at (axis cs: 12.05, -24)
  	            {\tiny{${m}_{c_i}$}\quad\quad\quad\;\tiny{$120$}\quad\quad\;\;\;\,\tiny{$200$}\quad\quad\;\;\;\,\tiny{$300$}};
  	            
                \draw [black] (axis cs:9.97,-21.35) -- (axis cs:10.78,-21.35);
                \draw [black, dashed] (axis cs:11.92,-21.35) -- (axis cs:12.73,-21.35);
                \draw [black, densely dash dot dot] (axis cs:13.80,-21.35) -- (axis cs:14.41,-21.35);
	        \end{axis} 
	    \end{tikzpicture}}
	\end{tabular}\vspace{0.1cm} 
	\captionsetup[subfigure]{oneside,margin={0.9cm,0cm}}
	\begin{tabular}{@{\hspace{-0.3cm}}c@{\hspace{0cm}}} \subfloat[]{\label{plot8b} \centering	
              \begin{tikzpicture}
      \begin{axis}[width=9cm, height=4.0cm,
	            grid style={line width=.3pt, draw=gray},
                x tick label style={/pgf/number format/.cd, set thousands separator={},fixed},
                ylabel style ={yshift=-0.1cm},
  	            xlabel={Local Iterations $\nu_l$},
  	            ylabel={Scale Factor $\phi$},
  	            label style={font=\scriptsize},
  	            grid=major,
  	            ymin = -40, ymax = 130, 
  	            xmin = 0.5, xmax = 15.5,
  	            xtick={1,2,3,4,5, 6,7,8,9,10, 11,12,13,14,15},
  	            xticklabels={1,5,30,60,90, 1,2,3,4,5, 1,2,3,4,5},
  	            tick label style={font=\scriptsize},
  	            legend style={draw=none, fill=gray!7!white, legend cell align=left, font=\tiny,
  	            at={(0.995, 0.985)}, anchor=north east, 
  	            {/tikz/every even column/.append style={column sep=0.1cm}}, 
  	            {/tikz/column 1/.append style={column sep=-0.2cm}},
  	            /tikz/column 4/.style={yshift=-0.025cm}, 
  	            /tikz/column 6/.style={yshift=-0.025cm}, 
  	            /tikz/column 8/.style={yshift=-0.025cm}, inner sep=0.5pt}, 
  	            legend columns=4]
                
                \addlegendimage{legend image with text=$\bar{\gamma}_{c_i}$}
                \addlegendentry{}; 
	            \addlegendimage{mark=diamond*, only marks, mark options={wind}}
	            \addlegendentry{$0.05m_{c_i}$}; 
                \addlegendimage{mark=otimes*, only marks, mark options={gas}}
	            \addlegendentry{$0.25m_{c_i}$}; 
	            \addlegendimage{mark=square*, only marks, mark options={biofuels}}
	            \addlegendentry{$0.5m_{c_i}$}; 
	            
                \addplot[mark=diamond*, mark size=1.7pt, mark options={wind}, black] 
  	            table [x={x}, y={y}] {./plot/plot8/plot8b/tie005_row120.txt}; 
  	            \addplot[mark=diamond*, mark size=1.7pt, mark options={wind}, black, dashed] 
  	            table [x={x}, y={y}] {./plot/plot8/plot8b/tie005_row200.txt}; 
  	            \addplot[mark=diamond*, mark size=1.7pt, mark options={wind}, black, densely dash dot dot] 
  	            table [x={x}, y={y}] {./plot/plot8/plot8b/tie005_row300.txt}; 
  	            \draw [thin] (axis cs:5.5,-40) -- (axis cs:5.5,140);  
  	            
	            \addplot[mark=otimes*, mark size=1.5pt, mark options={gas}, black] 
  	            table [x={x}, y={y}] {./plot/plot8/plot8b/tie025_row120.txt}; 
  	           	\addplot[mark=otimes*, mark size=1.5pt, mark options={gas}, black, dashed] 
  	            table [x={x}, y={y}] {./plot/plot8/plot8b/tie025_row200.txt};
  	            \addplot[mark=otimes*, mark size=1.7pt, mark options={gas}, black, densely dash dot dot] 
  	            table [x={x}, y={y}] {./plot/plot8/plot8b/tie025_row300.txt};
  	            \draw [thin] (axis cs:10.5,-20) -- (axis cs:10.5,120);
  	            
  	            \addplot[mark=square*, mark size=1.5pt, mark options={biofuels}, black]
  	            table [x={x}, y={y}] {./plot/plot8/plot8b/tie05_row120.txt}; 
  	           	\addplot[mark=square*, mark size=1.5pt, mark options={biofuels}, black, dashed] 
  	            table [x={x}, y={y}] {./plot/plot8/plot8a/tie05_row200.txt};  	            
  	            \addplot[mark=square*, mark size=1.7pt, mark options={biofuels}, black, densely dash dot dot] 
  	            table [x={x}, y={y}] {./plot/plot8/plot8b/tie05_row300.txt};
  	            
  	            \node[fill=gray!7!white, inner sep=2.3pt] at (axis cs: 12.05, -24)
  	            {\tiny{${m}_{c_i}$}\quad\quad\quad\;\tiny{$120$}\quad\quad\;\;\;\,\tiny{$200$}\quad\quad\;\;\;\,\tiny{$300$}};
  	            
                \draw [black] (axis cs:9.97,-21.35) -- (axis cs:10.78,-21.35);
                \draw [black, dashed] (axis cs:11.92,-21.35) -- (axis cs:12.73,-21.35);
                \draw [black, densely dash dot dot] (axis cs:13.80,-21.35) -- (axis cs:14.41,-21.35);
	        \end{axis} 
	    \end{tikzpicture}}
	\end{tabular}
	\caption{The median values of the scale factor $\phi$ depending on the different local iterations $\nu_l$, where the number of global iterations is equal to $\nu_g = 1$ for $\bar{\lambda}_{c_i} = 6m_{c_i}$ (subfigure a), and $\bar{\lambda}_{c_i} = 26m_{c_i}$ (subfigure b) for distributed system with $s = 2$, $m_{c_i} = \{120, 200, 300\}$, $\bar{\gamma}_{c_i} = \{0.05m_{c_i}, 0.25m_{c_i}, 0.5m_{c_i}\}$, $v_i = 10^{-8}$, $i = 1,\dots,n_{c_i}$, $v_i = 10^{-1}$, $i = n_{c_i} + 1,\dots,m_{c_i}$.}
	\label{plot8}
\end{figure}
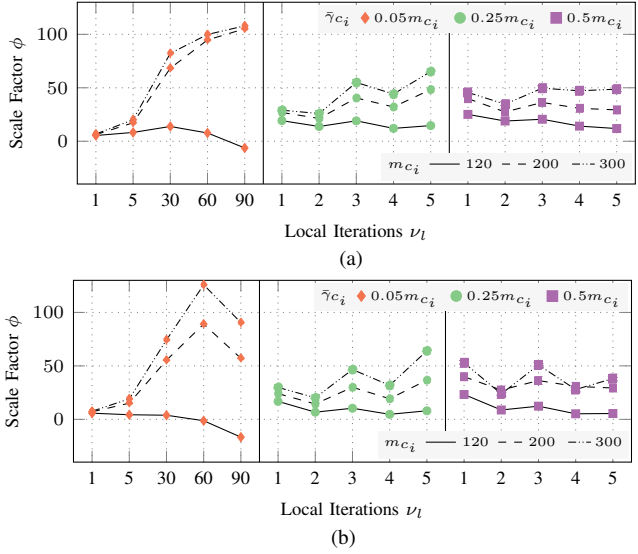

For experiments, we analyse models with $s = 2$ and $n_{c_i} = 100$, with different numbers of rows $m_{c_i} = \{120, 200, 300\}$. We compare the performance of the synchronous GBP and AGBP algorithm for the different number of expected internal edges equal to $\bar{\lambda}_{c_i} = 6m_{c_i}$ and $\bar{\lambda}_{c_i} = 26m_{c_i}$, defined by the matrix $\mathbf{H}_{c_i} \in \mathbb{R}^{m_{c_i} \times n_{c_i}}$. Furthermore, we observe the different number of expected tie edges $\bar{\gamma}_{c_i} = \{0.05m_{c_i}, 0.25m_{c_i}, 0.5m_{c_i}\}$ defined by the matrix $\mathbf{H}_{c_i,c_j} \in \mathbb{R}^{m_{c_i} \times n_{c_i}}$. Finally, we set the variances to the value of $v_i = 10^{-8}$, $i = 1,\dots,n_{c_i}$, for independent equations and $v_i = 10^{-1}$, $i = n_{c_i} + 1,\dots,m_{c_i}$, for dependent equations. In this setup, the synchronous GBP algorithm converges, whereas the AGBP algorithm converges only for specific global and local iteration schemes. In particular, for the case where the number of tie edges ${\gamma}_{c_i}$ increases while the number of internal edges ${\lambda}_{c_i}$ is small, reduced number of local iterations $\nu_l$ is needed to ensure convergence of the AGBP. \figurename~\ref{plot8} shows the median values of the scale factor $\phi$ for specific local and global iteration schemes under which the AGBP algorithm converges. Unlike systems with square matrices, the scale factor $\phi$ decreases as the number of internal edges $\lambda_{c_i}$ increases. Furthermore, the maximum values of the scale factor $\phi$ have smaller values compared to the case with square matrices. However, condition $\nu_s(\nu_g + \nu_l) < \nu$ is satisfied for a large number of local and global iteration schemes.
  
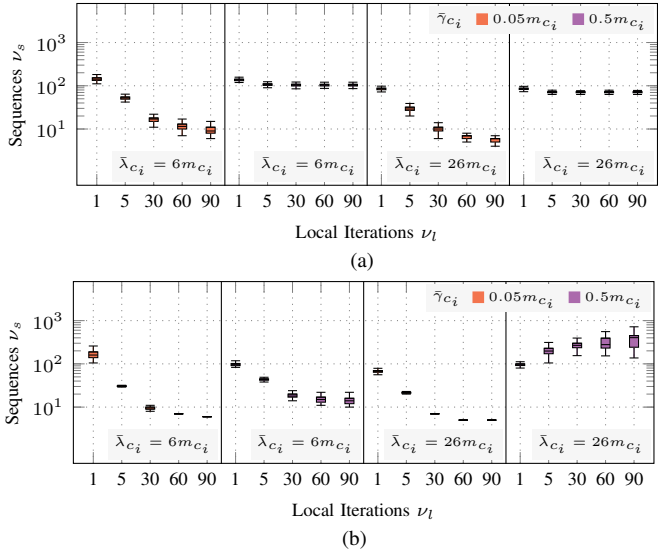
\begin{figure}[ht]
	\centering \captionsetup[subfigure]{oneside,margin={1.0cm,0cm}}
	\begin{tabular}{@{\hspace{-0.2cm}}c@{}} 
	\subfloat[]{\label{plot9a} \centering	
    \begin{tikzpicture}
	   \begin{axis} [box plot width=0.6mm, ymode=log, width=9.3cm, height=4.0cm,
            grid style={line width=.3pt, draw=gray},
	        xlabel style={align=center},
    	    xlabel={Local Iterations $\nu_l$},
            ylabel={Sequences $\nu_s$},
            ylabel style ={yshift=-0.1cm},
  	        grid=major,   		
  	        xmin=0.3, xmax=20.7, 
  	        ymin = 0.5, ymax = 8000,
            ytick={10,100,1000},
  	        xtick={1,2,3,4,5, 6,7,8,9,10, 11,12,13,14,15, 16,17,18,19,20},
            xticklabels={1,5,30,60,90, 1,5,30,60,90, 1,5,30,60,90, 1,5,30,60,90}, 
  	        tick label style={font=\scriptsize}, label style={font=\scriptsize},
  	        legend style={draw=none, fill=gray!7!white, legend cell align=left, font=\tiny,
  	        at={(0.995, 0.985)}, anchor=north east, 
  	        {/tikz/every even column/.append style={column sep=0.1cm}}, 
  	        {/tikz/column 1/.append style={column sep=-0.15cm}},
  	        /tikz/column 4/.style={yshift=-0.025cm}, 
  	        /tikz/column 6/.style={yshift=-0.025cm}, 
  	        inner sep=0.5pt}, legend columns=4]
                
            \addlegendimage{legend image with text=$\bar{\gamma}_{c_i}$}
            \addlegendentry{}; 
	        \addlegendimage{mark=square*, only marks, mark options={wind}}
	        \addlegendentry{$0.05m_{c_i}$}; 
	        \addlegendimage{mark=square*, only marks, mark options={biofuels}}
	        \addlegendentry{$0.5m_{c_i}$}; 
	
	        \boxplot [forget plot, fill=wind, box plot whisker bottom index=1,
            box plot whisker top index=5, box plot box bottom index=2,
            box plot box top index=4, box plot median index=3] 
            {./plot/plot9/plot9a/tie005_area6_row120.txt};
            \draw [thin] (axis cs:5.5,0.1) -- (axis cs:5.5,8000);                
            \node[fill=gray!7!white, inner sep=2.3pt, anchor=south east] at (axis cs: 5.5 -0.1, 0.5+0.1) {\tiny{$\bar{\lambda}_{c_i} = 6m_{c_i}$}};

            \boxplot [forget plot, fill=biofuels, box plot whisker bottom index=1,
            box plot whisker top index=5, box plot box bottom index=2,
            box plot box top index=4, box plot median index=3] 
            {./plot/plot9/plot9a/tie05_area6_row120.txt};
            \draw [thin] (axis cs:10.5,0.1) -- (axis cs:10.5,8000);
            \node[fill=gray!7!white, inner sep=2.3pt, anchor=south east] at (axis cs: 10.5 -0.1, 0.5+0.1) {\tiny{$\bar{\lambda}_{c_i} = 6m_{c_i}$}};
                
            \boxplot [forget plot, fill=wind, box plot whisker bottom index=1,
            box plot whisker top index=5, box plot box bottom index=2,
            box plot box top index=4, box plot median index=3] 
            {./plot/plot9/plot9a/tie005_area26_row120.txt};
            \draw [thin] (axis cs:15.5,0.1) -- (axis cs:15.5,8000);
            \node[fill=gray!7!white, inner sep=2.3pt, anchor=south east] at (axis cs: 15.5 -0.1, 0.5+0.1) {\tiny{$\bar{\lambda}_{c_i} = 26m_{c_i}$}};
                
            \boxplot [forget plot, fill=biofuels, box plot whisker bottom index=1,
            box plot whisker top index=5, box plot box bottom index=2,
            box plot box top index=4, box plot median index=3] 
            {./plot/plot9/plot9a/tie05_area26_row120.txt};
            \node[fill=gray!7!white, inner sep=2.3pt, anchor=south east] at (axis cs: 20.7 -0.1, 0.5+0.1) {\tiny{$\bar{\lambda}_{c_i} = 26m_{c_i}$}};
                
	        \end{axis}
	    \end{tikzpicture}}
	\end{tabular}\vspace{0.1cm}
	\captionsetup[subfigure]{oneside,margin={0.9cm,0cm}}
	\begin{tabular}{@{\hspace{-0.3cm}}c@{\hspace{0cm}}} \subfloat[]{\label{plot9b} \centering	
    \begin{tikzpicture}
   	   \begin{axis} [box plot width=0.6mm, ymode=log, width=9.3cm, height=4.0cm,
            grid style={line width=.3pt, draw=gray},
	        xlabel style={align=center},
    	    xlabel={Local Iterations $\nu_l$},
            ylabel={Sequences $\nu_s$},
            ylabel style ={yshift=-0.1cm},
  	        grid=major,   		
  	        xmin=0.3, xmax=20.7, 
  	        ymin = 0.5, ymax = 8000,
            ytick={10,100,1000},
  	        xtick={1,2,3,4,5, 6,7,8,9,10, 11,12,13,14,15, 16,17,18,19,20},
            xticklabels={1,5,30,60,90, 1,5,30,60,90, 1,5,30,60,90, 1,5,30,60,90}, 
  	        tick label style={font=\scriptsize}, label style={font=\scriptsize},
  	        legend style={draw=none, fill=gray!7!white, legend cell align=left, font=\tiny,
  	        at={(0.995, 0.985)}, anchor=north east, 
  	        {/tikz/every even column/.append style={column sep=0.1cm}}, 
  	        {/tikz/column 1/.append style={column sep=-0.15cm}},
  	        /tikz/column 4/.style={yshift=-0.025cm}, 
  	        /tikz/column 6/.style={yshift=-0.025cm}, 
  	        inner sep=0.5pt}, legend columns=4]
                
            \addlegendimage{legend image with text=$\bar{\gamma}_{c_i}$}
            \addlegendentry{}; 
	        \addlegendimage{mark=square*, only marks, mark options={wind}}
	        \addlegendentry{$0.05m_{c_i}$}; 
	        \addlegendimage{mark=square*, only marks, mark options={biofuels}}
	        \addlegendentry{$0.5m_{c_i}$}; 
	
	        \boxplot [forget plot, fill=wind, box plot whisker bottom index=1,
            box plot whisker top index=5, box plot box bottom index=2,
            box plot box top index=4, box plot median index=3] 
            {./plot/plot9/plot9b/tie005_area6_row300.txt};
            \draw [thin] (axis cs:5.5,0.1) -- (axis cs:5.5,8000);                
            \node[fill=gray!7!white, inner sep=2.3pt, anchor=south east] at (axis cs: 5.5 -0.1, 0.5+0.1) {\tiny{$\bar{\lambda}_{c_i} = 6m_{c_i}$}};

            \boxplot [forget plot, fill=biofuels, box plot whisker bottom index=1,
            box plot whisker top index=5, box plot box bottom index=2,
            box plot box top index=4, box plot median index=3] 
            {./plot/plot9/plot9b/tie05_area6_row300.txt};
            \draw [thin] (axis cs:10.5,0.1) -- (axis cs:10.5,8000);
            \node[fill=gray!7!white, inner sep=2.3pt, anchor=south east] at (axis cs: 10.5 -0.1, 0.5+0.1) {\tiny{$\bar{\lambda}_{c_i} = 6m_{c_i}$}};
                
            \boxplot [forget plot, fill=wind, box plot whisker bottom index=1,
            box plot whisker top index=5, box plot box bottom index=2,
            box plot box top index=4, box plot median index=3] 
            {./plot/plot9/plot9b/tie005_area26_row300.txt};
            \draw [thin] (axis cs:15.5,0.1) -- (axis cs:15.5,8000);
            \node[fill=gray!7!white, inner sep=2.3pt, anchor=south east] at (axis cs: 15.5 -0.1, 0.5+0.1) {\tiny{$\bar{\lambda}_{c_i} = 26m_{c_i}$}};
                
            \boxplot [forget plot, fill=biofuels, box plot whisker bottom index=1,
            box plot whisker top index=5, box plot box bottom index=2,
            box plot box top index=4, box plot median index=3] 
            {./plot/plot9/plot9b/tie05_area26_row300.txt};
            \node[fill=gray!7!white, inner sep=2.3pt, anchor=south east] at (axis cs: 20.7 -0.1, 0.5+0.1) {\tiny{$\bar{\lambda}_{c_i} = 26m_{c_i}$}};
                
	        \end{axis}
	    \end{tikzpicture}}
	\end{tabular}
	\caption{The number of sequences of the AGBP algorithm, where the number of global iterations is equal to $\nu_g = 1$, for ${m}_{c_i} = 120$ (subfigure a), and ${m}_{c_i} = 300$ (subfigure b) for distributed system with $s = 2$, $\bar{\gamma}_{c_i} = \{0.05m_{c_i}, 0.5m_{c_i}\}$, $v_i = 10^{-1}$, $i = 1,\dots,m_{c_i}$, with randomised damping parameters equal to $\zeta = 0.9$, $p = 0.9$ for $\bar{\gamma}_{c_i} = 0.05m_{c_i}$, and $p = 0.7$ for $\bar{\gamma}_{c_i} = 0.5m_{c_i}$ (see Appendix B for details).}
	\label{plot9}
\end{figure}

Next, we reduce the difference between the variances associated with independent and dependent equations, resulting in a significantly reduced probability of convergence for both, synchronous GBP and AGBP algorithms. However, unlike with the synchronous GBP, the use of damping (see Appendix D for details) leads to convergence of the AGBP. More precisely, combining the damping during local iterations leads to a significant reduction in the spectral radius of the resulting matrix, which affects the evolution of the mean values. In addition, the AGBP algorithm converges for any global and local iteration scheme. Adding new rows  has a positive impact on the performance of the AGBP algorithm, resulting in a small number of sequences $\nu_s$ required to reach convergence for the majority of the analysed systems, as depicted in  \figurename~\ref{plot9}. Performance degrades only for a few cases with a large number of tie edges $\gamma_{c_i}$.

Finally, we test scenarios with a larger number of clusters (i.e., $s = 6$ and $s = 12$), and inflated factor graphs, where inflated means increased number of internal edges, tie edges, variable, and factor nodes by factor of three. In these scenarios, we observe a negligible increase in the number of sequences $\nu_s$.

\subsection{Linear Models with Rectangular Matrices in Dynamic Framework}
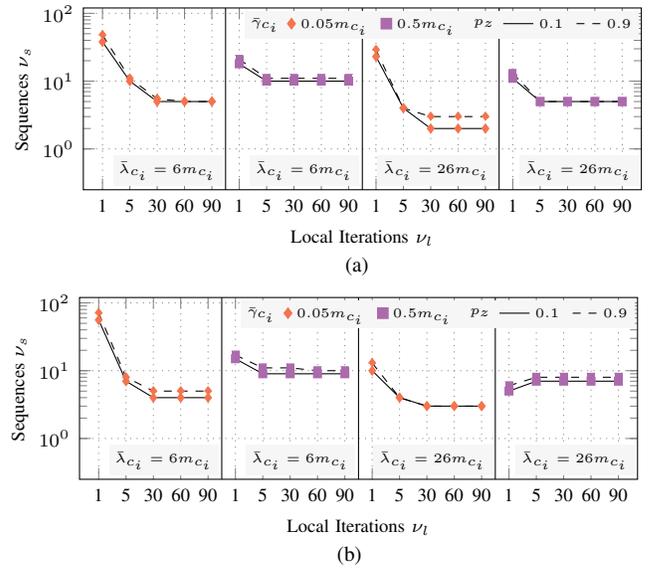
\begin{figure}[ht]
	\centering \captionsetup[subfigure]{oneside,margin={1.0cm,0cm}}
	\begin{tabular}{@{\hspace{-0.2cm}}c@{}} 
	\subfloat[]{\label{plot10a} \centering	
        \begin{tikzpicture}
            \begin{axis}[width=9cm, height=4.0cm, ymode=log,
	            grid style={line width=.3pt, draw=gray},
                x tick label style={/pgf/number format/.cd, set thousands separator={},fixed},
                ylabel style ={yshift=-0.1cm},
  	            xlabel={Local Iterations $\nu_l$},
  	            ylabel={Sequences $\nu_s$},
  	            label style={font=\scriptsize},
  	            grid=major,
  	            ymin = 0.25, ymax = 120, 
  	            xmin=0.3, xmax=20.7,
  	            ytick={1, 10, 100},
  	            xtick={1,2,3,4,5, 6,7,8,9,10, 11,12,13,14,15, 16,17,18,19,20},
  	            xticklabels={1,5,30,60,90, 1,5,30,60,90, 1,5,30,60,90, 1,5,30,60,90},
  	            tick label style={font=\scriptsize},
    	        legend style={draw=none, fill=gray!7!white, legend cell align=left, font=\tiny,
  	            at={(0.995, 0.985)}, anchor=north east, 
  	            {/tikz/every even column/.append style={column sep=0.1cm}}, 
  	            {/tikz/column 1/.append style={column sep=-0.2cm}},
  	            /tikz/column 4/.style={yshift=-0.025cm}, 
  	            /tikz/column 6/.style={yshift=-0.025cm}, 
  	            {/tikz/column 7/.append style={column sep=-0.2cm}},
  	            /tikz/column 10/.style={yshift=-0.025cm}, 
  	            /tikz/column 12/.style={yshift=-0.025cm}, 
  	            {/tikz/column 11/.append style={column sep=-0.07cm}},
  	            inner sep=0.5pt}, legend columns=10]
                
                \addlegendimage{legend image with text=$\bar{\gamma}_{c_i}$}
                \addlegendentry{}; 
	            \addlegendimage{mark=diamond*, only marks, mark options={wind}}
	            \addlegendentry{$0.05m_{c_i}$}; 
	            \addlegendimage{mark=square*, only marks, mark options={biofuels}}
	            \addlegendentry{$0.5m_{c_i}$}; 
	            \addlegendimage{legend image with text=$p_z$}
	            \addlegendentry{};
	            \addlegendimage{legend image code/.code={\draw[black] plot coordinates {(0cm,0cm) (0.4cm,0cm)};}}
	            \addlegendentry{$0.1$}; 
	            \addlegendimage{legend image code/.code={\draw[black, dashed] plot coordinates {(0cm,0cm) (0.4cm,0cm)};}}
	            \addlegendentry{$0.9$}; 
	            
                \addplot[mark=diamond*, mark size=1.7pt, mark options={wind}, black] 
  	            table [x={x}, y={y}] {./plot/plot10/plot10a/tie005_area6_dyn01.txt}; 
  	            \addplot[mark=diamond*, mark size=1.7pt, mark options={wind}, black, dashed] 
  	            table [x={x}, y={y}] {./plot/plot10/plot10a/tie005_area6_dyn09.txt}; 
  	            \draw [thin] (axis cs:5.5,0.2) -- (axis cs:5.5,120);  	            
  	            \node[fill=gray!7!white, inner sep=2.3pt, anchor=south east] at (axis cs: 5.5 -0.1, 0.23+0.05) {\tiny{$\bar{\lambda}_{c_i} = 6m_{c_i}$}};
  
	            \addplot[mark=square*, mark size=1.5pt, mark options={biofuels}, black] 
  	            table [x={x}, y={y}] {./plot/plot10/plot10a/tie05_area6_dyn01.txt}; 
  	           	\addplot[mark=square*, mark size=1.5pt, mark options={biofuels}, black, dashed] 
  	            table [x={x}, y={y}] {./plot/plot10/plot10a/tie05_area6_dyn09.txt}; 
  	            \draw [thin] (axis cs:10.5,0.2) -- (axis cs:10.5,120);
                \node[fill=gray!7!white, inner sep=2.3pt, anchor=south east] at (axis cs: 10.5 -0.1, 0.23+0.05) {\tiny{$\bar{\lambda}_{c_i} = 6m_{c_i}$}};  	            
  	            
                \addplot[mark=diamond*, mark size=1.7pt, mark options={wind}, black] 
  	            table [x={x}, y={y}] {./plot/plot10/plot10a/tie005_area26_dyn01.txt}; 
  	            \addplot[mark=diamond*, mark size=1.7pt, mark options={wind}, black, dashed] 
  	            table [x={x}, y={y}] {./plot/plot10/plot10a/tie005_area26_dyn09.txt}; 
  	            \draw [thin] (axis cs:15.5,0.2) -- (axis cs:15.5,120);
                \node[fill=gray!7!white, inner sep=2.3pt, anchor=south east] at (axis cs: 15.5 -0.1, 0.23+0.05) {\tiny{$\bar{\lambda}_{c_i} = 26m_{c_i}$}};

                \addplot[mark=square*, mark size=1.5pt, mark options={biofuels}, black] 
  	            table [x={x}, y={y}] {./plot/plot10/plot10a/tie05_area26_dyn01.txt}; 
  	           	\addplot[mark=square*, mark size=1.5pt, mark options={biofuels}, black, dashed] 
  	            table [x={x}, y={y}] {./plot/plot10/plot10a/tie05_area26_dyn09.txt};   	         
  	            \node[fill=gray!7!white, inner sep=2.3pt, anchor=south east] at (axis cs: 20.7 -0.1, 0.23+0.05) {\tiny{$\bar{\lambda}_{c_i} = 26m_{c_i}$}};
	        \end{axis} 
	    \end{tikzpicture}}
	\end{tabular}\vspace{0.1cm}
	\captionsetup[subfigure]{oneside,margin={0.9cm,0cm}}
	\begin{tabular}{@{\hspace{-0.3cm}}c@{\hspace{0cm}}} \subfloat[]{\label{plot10b} \centering	
        \begin{tikzpicture}
            \begin{axis}[width=9cm, height=4.0cm, ymode=log,
	            grid style={line width=.3pt, draw=gray},
                x tick label style={/pgf/number format/.cd, set thousands separator={},fixed},
                ylabel style ={yshift=-0.1cm},
  	            xlabel={Local Iterations $\nu_l$},
  	            ylabel={Sequences $\nu_s$},
  	            label style={font=\scriptsize},
  	            grid=major,
  	            ymin = 0.25, ymax = 120, 
  	            xmin=0.3, xmax=20.7,
  	            ytick={1, 10, 100},
  	            xtick={1,2,3,4,5, 6,7,8,9,10, 11,12,13,14,15, 16,17,18,19,20},
  	            xticklabels={1,5,30,60,90, 1,5,30,60,90, 1,5,30,60,90, 1,5,30,60,90},
  	            tick label style={font=\scriptsize},
    	        legend style={draw=none, fill=gray!7!white, legend cell align=left, font=\tiny,
  	            at={(0.995, 0.985)}, anchor=north east, 
  	            {/tikz/every even column/.append style={column sep=0.1cm}}, 
  	            {/tikz/column 1/.append style={column sep=-0.2cm}},
  	            /tikz/column 4/.style={yshift=-0.025cm}, 
  	            /tikz/column 6/.style={yshift=-0.025cm}, 
  	            {/tikz/column 7/.append style={column sep=-0.2cm}},
  	            /tikz/column 10/.style={yshift=-0.025cm}, 
  	            /tikz/column 12/.style={yshift=-0.025cm}, 
  	            {/tikz/column 11/.append style={column sep=-0.07cm}},
  	            inner sep=0.5pt}, legend columns=10]
                
                \addlegendimage{legend image with text=$\bar{\gamma}_{c_i}$}
                \addlegendentry{}; 
	            \addlegendimage{mark=diamond*, only marks, mark options={wind}}
	            \addlegendentry{$0.05m_{c_i}$}; 
	            \addlegendimage{mark=square*, only marks, mark options={biofuels}}
	            \addlegendentry{$0.5m_{c_i}$}; 
	            \addlegendimage{legend image with text=$p_z$}
	            \addlegendentry{};
	            \addlegendimage{legend image code/.code={\draw[black] plot coordinates {(0cm,0cm) (0.4cm,0cm)};}}
	            \addlegendentry{$0.1$}; 
	            \addlegendimage{legend image code/.code={\draw[black, dashed] plot coordinates {(0cm,0cm) (0.4cm,0cm)};}}
	            \addlegendentry{$0.9$}; 
	            
                \addplot[mark=diamond*, mark size=1.7pt, mark options={wind}, black] 
  	            table [x={x}, y={y}] {./plot/plot10/plot10b/tie005_area6_dyn01.txt}; 
  	            \addplot[mark=diamond*, mark size=1.7pt, mark options={wind}, black, dashed] 
  	            table [x={x}, y={y}] {./plot/plot10/plot10b/tie005_area6_dyn09.txt}; 
  	            \draw [thin] (axis cs:5.5,0.2) -- (axis cs:5.5,120);  	            
  	            \node[fill=gray!7!white, inner sep=2.3pt, anchor=south east] at (axis cs: 5.5 -0.1, 0.23+0.05) {\tiny{$\bar{\lambda}_{c_i} = 6m_{c_i}$}};
  
	            \addplot[mark=square*, mark size=1.5pt, mark options={biofuels}, black] 
  	            table [x={x}, y={y}] {./plot/plot10/plot10b/tie05_area6_dyn01.txt}; 
  	           	\addplot[mark=square*, mark size=1.5pt, mark options={biofuels}, black, dashed] 
  	            table [x={x}, y={y}] {./plot/plot10/plot10b/tie05_area6_dyn09.txt}; 
  	            \draw [thin] (axis cs:10.5,0.2) -- (axis cs:10.5,120);
                \node[fill=gray!7!white, inner sep=2.3pt, anchor=south east] at (axis cs: 10.5 -0.1, 0.23+0.05) {\tiny{$\bar{\lambda}_{c_i} = 6m_{c_i}$}};  	            
  	            
                \addplot[mark=diamond*, mark size=1.7pt, mark options={wind}, black] 
  	            table [x={x}, y={y}] {./plot/plot10/plot10b/tie005_area26_dyn01.txt}; 
  	            \addplot[mark=diamond*, mark size=1.7pt, mark options={wind}, black, dashed] 
  	            table [x={x}, y={y}] {./plot/plot10/plot10b/tie005_area26_dyn09.txt}; 
  	            \draw [thin] (axis cs:15.5,0.2) -- (axis cs:15.5,120);
                \node[fill=gray!7!white, inner sep=2.3pt, anchor=south east] at (axis cs: 15.5 -0.1, 0.23+0.05) {\tiny{$\bar{\lambda}_{c_i} = 26m_{c_i}$}};

                \addplot[mark=square*, mark size=1.5pt, mark options={biofuels}, black] 
  	            table [x={x}, y={y}] {./plot/plot10/plot10b/tie05_area26_dyn01.txt}; 
  	           	\addplot[mark=square*, mark size=1.5pt, mark options={biofuels}, black, dashed] 
  	            table [x={x}, y={y}] {./plot/plot10/plot10b/tie05_area26_dyn09.txt};   	         
  	            \node[fill=gray!7!white, inner sep=2.3pt, anchor=south east] at (axis cs: 20.7 -0.1, 0.23+0.05) {\tiny{$\bar{\lambda}_{c_i} = 26m_{c_i}$}};
	        \end{axis} 
	    \end{tikzpicture}}
	\end{tabular}
	\caption{The median values of the number of sequences of the AGBP algorithm, where the number of global iterations is equal to $\nu_g = 1$, for ${m}_{c_i} = 120$ (subfigure a), and ${m}_{c_i} = 300$ (subfigure b), where we change each element of the vector of observation values $\mathbf{z}$ with probability $p_z =0.1$ and $p_z =0.9$ for distributed system with $s = 2$, $\bar{\gamma}_{c_i} = \{0.05m_{c_i}, 0.5m_{c_i}\}$, $v_i = 10^{-1}$, $i = 1,\dots,m_{c_i}$, with randomised damping parameters equal to $\zeta = 0.9$, $p = 0.9$ for $\bar{\gamma}_{c_i} = 0.05m_{c_i}$, and $p = 0.7$ for $\bar{\gamma}_{c_i} = 0.5m_{c_i}$ (see Appendix B for details).}
	\label{plot10}
\end{figure}

We extend the analysis in Section \ref{rectangular} by allowing a dynamic change of values in vector $\mathbf{z}$, thus simulating more realistic case where new observations arrive over time. \figurename~\ref{plot10} illustrates the number of sequences $\nu_s$ needed for the convergence of the AGBP algorithm. Following the algorithm convergence, we randomly change the observation values from the vector $\mathbf{z}$ according to a uniform distribution with probability $p_z = 0.1$ and $p_z = 0.9$, and continue the iterative process. In the dynamic framework, the AGBP algorithm converges to a new fixed point in few sequences $\nu_s$, as illustrated in \figurename~\ref{plot10}. Furthermore, there is no significant impact of the small $p_z = 0.1$ and large $p_z = 0.9$ probabilities on the convergence trend in the observational values.

\begin{figure}[ht]
	\centering \captionsetup[subfigure]{oneside,margin={1.0cm,0cm}}
	\begin{tabular}{@{\hspace{-0.2cm}}c@{}} 
	\subfloat[]{\label{plot11a} \centering	
        \begin{tikzpicture}
            \begin{axis}[width=9cm, height=4.0cm, ymode=log,
	            grid style={line width=.3pt, draw=gray},
                x tick label style={/pgf/number format/.cd, set thousands separator={},fixed},
                ylabel style ={yshift=-0.1cm},
  	            xlabel={Local Iterations $\nu_l$},
  	            ylabel={Sequences $\nu_s$},
  	            label style={font=\scriptsize},
  	            grid=major,
  	            ymin = 0.25, ymax = 120, 
  	            xmin=0.3, xmax=20.7,
  	            ytick={1, 10, 100},
  	            xtick={1,2,3,4,5, 6,7,8,9,10, 11,12,13,14,15, 16,17,18,19,20},
  	            xticklabels={1,5,30,60,90, 1,5,30,60,90, 1,5,30,60,90, 1,5,30,60,90},
  	            tick label style={font=\scriptsize},
    	        legend style={draw=none, fill=gray!7!white, legend cell align=left, font=\tiny,
  	            at={(0.995, 0.985)}, anchor=north east, 
  	            {/tikz/every even column/.append style={column sep=0.1cm}}, 
  	            {/tikz/column 1/.append style={column sep=-0.2cm}},
  	            /tikz/column 4/.style={yshift=-0.025cm}, 
  	            /tikz/column 6/.style={yshift=-0.025cm}, 
  	            {/tikz/column 7/.append style={column sep=-0.2cm}},
  	            /tikz/column 10/.style={yshift=-0.025cm}, 
  	            /tikz/column 12/.style={yshift=-0.025cm}, 
  	            {/tikz/column 11/.append style={column sep=-0.07cm}},
  	            inner sep=0.5pt}, legend columns=10]
                
                \addlegendimage{legend image with text=$\bar{\gamma}_{c_i}$}
                \addlegendentry{}; 
	            \addlegendimage{mark=diamond*, only marks, mark options={wind}}
	            \addlegendentry{$0.05m_{c_i}$}; 
	            \addlegendimage{mark=square*, only marks, mark options={biofuels}}
	            \addlegendentry{$0.5m_{c_i}$}; 
	            \addlegendimage{legend image with text=$p_z$}
	            \addlegendentry{};
	            \addlegendimage{legend image code/.code={\draw[black] plot coordinates {(0cm,0cm) (0.4cm,0cm)};}}
	            \addlegendentry{$0.1$}; 
	            \addlegendimage{legend image code/.code={\draw[black, dashed] plot coordinates {(0cm,0cm) (0.4cm,0cm)};}}
	            \addlegendentry{$0.9$}; 
	            
                \addplot[mark=diamond*, mark size=1.7pt, mark options={wind}, black] 
  	            table [x={x}, y={y}] {./plot/plot11/plot11a/tie005_area6_dyn01.txt}; 
  	            \addplot[mark=diamond*, mark size=1.7pt, mark options={wind}, black, dashed] 
  	            table [x={x}, y={y}] {./plot/plot11/plot11a/tie005_area6_dyn09.txt}; 
  	            \draw [thin] (axis cs:5.5,0.2) -- (axis cs:5.5,120);  	            
  	            \node[fill=gray!7!white, inner sep=2.3pt, anchor=south east] at (axis cs: 5.5 -0.1, 0.23+0.05) {\tiny{$\bar{\lambda}_{c_i} = 6m_{c_i}$}};
  
	            \addplot[mark=square*, mark size=1.5pt, mark options={biofuels}, black] 
  	            table [x={x}, y={y}] {./plot/plot11/plot11a/tie05_area6_dyn01.txt}; 
  	           	\addplot[mark=square*, mark size=1.5pt, mark options={biofuels}, black, dashed] 
  	            table [x={x}, y={y}] {./plot/plot11/plot11a/tie05_area6_dyn09.txt}; 
  	            \draw [thin] (axis cs:10.5,0.2) -- (axis cs:10.5,120);
                \node[fill=gray!7!white, inner sep=2.3pt, anchor=south east] at (axis cs: 10.5 -0.1, 0.23+0.05) {\tiny{$\bar{\lambda}_{c_i} = 6m_{c_i}$}};  	            
  	            
                \addplot[mark=diamond*, mark size=1.7pt, mark options={wind}, black] 
  	            table [x={x}, y={y}] {./plot/plot11/plot11a/tie005_area26_dyn01.txt}; 
  	            \addplot[mark=diamond*, mark size=1.7pt, mark options={wind}, black, dashed] 
  	            table [x={x}, y={y}] {./plot/plot11/plot11a/tie005_area26_dyn09.txt}; 
  	            \draw [thin] (axis cs:15.5,0.2) -- (axis cs:15.5,120);
                \node[fill=gray!7!white, inner sep=2.3pt, anchor=south east] at (axis cs: 15.5 -0.1, 0.23+0.05) {\tiny{$\bar{\lambda}_{c_i} = 26m_{c_i}$}};

                \addplot[mark=square*, mark size=1.5pt, mark options={biofuels}, black] 
  	            table [x={x}, y={y}] {./plot/plot11/plot11a/tie05_area26_dyn01.txt}; 
  	           	\addplot[mark=square*, mark size=1.5pt, mark options={biofuels}, black, dashed] 
  	            table [x={x}, y={y}] {./plot/plot11/plot11a/tie05_area26_dyn09.txt};   	         
  	            \node[fill=gray!7!white, inner sep=2.3pt, anchor=south east] at (axis cs: 20.7 -0.1, 0.23+0.05) {\tiny{$\bar{\lambda}_{c_i} = 26m_{c_i}$}};
	        \end{axis} 
	    \end{tikzpicture}}
	\end{tabular}\vspace{0.1cm}
	\captionsetup[subfigure]{oneside,margin={0.9cm,0cm}}
	\begin{tabular}{@{\hspace{-0.3cm}}c@{\hspace{0cm}}} \subfloat[]{\label{plot11b} \centering	
        \begin{tikzpicture}
            \begin{axis}[width=9cm, height=4.0cm, ymode=log,
	            grid style={line width=.3pt, draw=gray},
                x tick label style={/pgf/number format/.cd, set thousands separator={},fixed},
                ylabel style ={yshift=-0.1cm},
  	            xlabel={Local Iterations $\nu_l$},
  	            ylabel={Sequences $\nu_s$},
  	            label style={font=\scriptsize},
  	            grid=major,
  	            ymin = 0.25, ymax = 120, 
  	            xmin=0.3, xmax=20.7,
  	            ytick={1, 10, 100},
  	            xtick={1,2,3,4,5, 6,7,8,9,10, 11,12,13,14,15, 16,17,18,19,20},
  	            xticklabels={1,5,30,60,90, 1,5,30,60,90, 1,5,30,60,90, 1,5,30,60,90},
  	            tick label style={font=\scriptsize},
    	        legend style={draw=none, fill=gray!7!white, legend cell align=left, font=\tiny,
  	            at={(0.995, 0.985)}, anchor=north east, 
  	            {/tikz/every even column/.append style={column sep=0.1cm}}, 
  	            {/tikz/column 1/.append style={column sep=-0.2cm}},
  	            /tikz/column 4/.style={yshift=-0.025cm}, 
  	            /tikz/column 6/.style={yshift=-0.025cm}, 
  	            {/tikz/column 7/.append style={column sep=-0.2cm}},
  	            /tikz/column 10/.style={yshift=-0.025cm}, 
  	            /tikz/column 12/.style={yshift=-0.025cm}, 
  	            {/tikz/column 11/.append style={column sep=-0.07cm}},
  	            inner sep=0.5pt}, legend columns=10]
                
                \addlegendimage{legend image with text=$\bar{\gamma}_{c_i}$}
                \addlegendentry{}; 
	            \addlegendimage{mark=diamond*, only marks, mark options={wind}}
	            \addlegendentry{$0.05m_{c_i}$}; 
	            \addlegendimage{mark=square*, only marks, mark options={biofuels}}
	            \addlegendentry{$0.5m_{c_i}$}; 
	            \addlegendimage{legend image with text=$p_z$}
	            \addlegendentry{};
	            \addlegendimage{legend image code/.code={\draw[black] plot coordinates {(0cm,0cm) (0.4cm,0cm)};}}
	            \addlegendentry{$0.1$}; 
	            \addlegendimage{legend image code/.code={\draw[black, dashed] plot coordinates {(0cm,0cm) (0.4cm,0cm)};}}
	            \addlegendentry{$0.9$}; 
	            
                \addplot[mark=diamond*, mark size=1.7pt, mark options={wind}, black] 
  	            table [x={x}, y={y}] {./plot/plot11/plot11b/tie005_area6_dyn01.txt}; 
  	            \addplot[mark=diamond*, mark size=1.7pt, mark options={wind}, black, dashed] 
  	            table [x={x}, y={y}] {./plot/plot11/plot11b/tie005_area6_dyn09.txt}; 
  	            \draw [thin] (axis cs:5.5,0.2) -- (axis cs:5.5,120);  	            
  	            \node[fill=gray!7!white, inner sep=2.3pt, anchor=south east] at (axis cs: 5.5 -0.1, 0.23+0.05) {\tiny{$\bar{\lambda}_{c_i} = 6m_{c_i}$}};
  
	            \addplot[mark=square*, mark size=1.5pt, mark options={biofuels}, black] 
  	            table [x={x}, y={y}] {./plot/plot11/plot11b/tie05_area6_dyn01.txt}; 
  	           	\addplot[mark=square*, mark size=1.5pt, mark options={biofuels}, black, dashed] 
  	            table [x={x}, y={y}] {./plot/plot11/plot11b/tie05_area6_dyn09.txt}; 
  	            \draw [thin] (axis cs:10.5,0.2) -- (axis cs:10.5,120);
                \node[fill=gray!7!white, inner sep=2.3pt, anchor=south east] at (axis cs: 10.5 -0.1, 0.23+0.05) {\tiny{$\bar{\lambda}_{c_i} = 6m_{c_i}$}};  	            
  	            
                \addplot[mark=diamond*, mark size=1.7pt, mark options={wind}, black] 
  	            table [x={x}, y={y}] {./plot/plot11/plot11b/tie005_area26_dyn01.txt}; 
  	            \addplot[mark=diamond*, mark size=1.7pt, mark options={wind}, black, dashed] 
  	            table [x={x}, y={y}] {./plot/plot11/plot11b/tie005_area26_dyn09.txt}; 
  	            \draw [thin] (axis cs:15.5,0.2) -- (axis cs:15.5,120);
                \node[fill=gray!7!white, inner sep=2.3pt, anchor=south east] at (axis cs: 15.5 -0.1, 0.23+0.05) {\tiny{$\bar{\lambda}_{c_i} = 26m_{c_i}$}};

                \addplot[mark=square*, mark size=1.5pt, mark options={biofuels}, black] 
  	            table [x={x}, y={y}] {./plot/plot11/plot11b/tie05_area26_dyn01.txt}; 
  	           	\addplot[mark=square*, mark size=1.5pt, mark options={biofuels}, black, dashed] 
  	            table [x={x}, y={y}] {./plot/plot11/plot11b/tie05_area26_dyn09.txt};   	         
  	            \node[fill=gray!7!white, inner sep=2.3pt, anchor=south east] at (axis cs: 20.7 -0.1, 0.23+0.05) {\tiny{$\bar{\lambda}_{c_i} = 26m_{c_i}$}};
	        \end{axis} 
	    \end{tikzpicture}}
	\end{tabular}
	\caption{The median values of the number of sequences of the AGBP algorithm, where the number of global iterations is equal to $\nu_g = 1$, for ${m}_{c_i} = 120$ (subfigure a), and ${m}_{c_i} = 300$ (subfigure b), where we change observation variance values $v_i$, $i = n_{c_i} + 1,\dots,m_{c_i}$ using logarithmic growth model \eqref{logarithmic} with probability $p_z =0.1$ and $p_z =0.9$ for distributed system with $s = 2$, $\bar{\gamma}_{c_i} = \{0.05m_{c_i}, 0.5m_{c_i}\}$, $v_i = 10^{-1}$, $i = 1,\dots,n_{c_i}$, with randomised damping parameters equal to $\zeta = 0.9$, $p = 0.9$ for $\bar{\gamma}_{c_i} = 0.05m_{c_i}$, and $p = 0.7$ for $\bar{\gamma}_{c_i} = 0.5m_{c_i}$ (see Appendix D for details).}
	\label{plot11}
\end{figure}
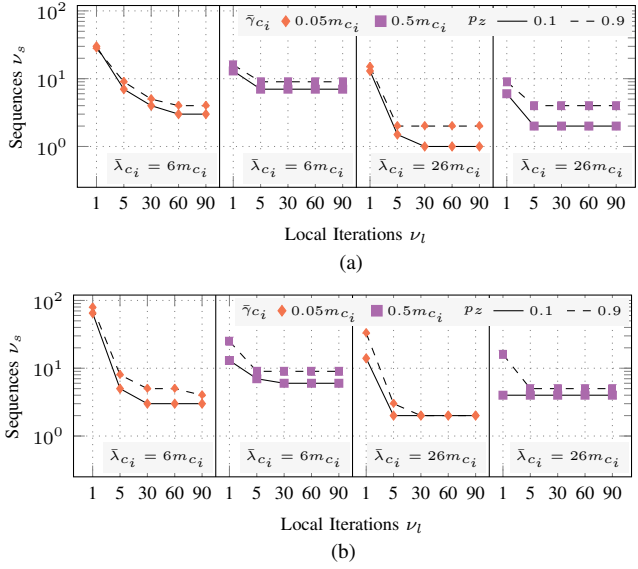

Further, we analyse the performance of the AGBP algorithm with the addition of a mechanism for adaptive increase of variance values. We increase the variance values of the dependent equations using the logarithmic growth model \eqref{logarithmic} with probability $p_z =0.1$ and $p_z =0.9$ for the same distributed system as analysed above. According to \figurename~\ref{plot11}, the AGBP algorithm reaches a new fixed point in a fewer sequences $\nu_s$, compared to the same scenarios in which we changed the observation values only (see \figurename~\ref{plot10}). 

The introduction of the dynamic AGBP algorithm has a positive impact on the number of iterations needed to converge compared to the uniform initialisation of messages presented in Sections \ref{symmetric} through \ref{rectangular}. The significant reduction in the number of iterations needed for convergence makes the dynamic AGBP algorithm a good candidate for large-scale, time-constrained systems where arrival of new data requires fast inference.

\section{Conclusions}
This paper analyses large-scale distributed systems by analysing maximum likelihood estimation using the GBP algorithm applied over clustered factor graphs. We show that the AGBP algorithm achieves a significant improvement in inference time compared to the synchronous GBP algorithm. Furthermore, we demonstrate the scalable properties of AGBP, making it suitable for large-scale inference in massive IoT networks. The proposed solution augmented with the damping technique achieves convergence where existing state-of-the-art methods fail. Due to the inherent properties of the GBP algorithm applied over the distributed architecture, where clusters exchange only ``beliefs" about specific state variables, the proposed framework ensures data privacy. Building upon the research initiated in this paper, where convergence analysis is based on the global view that exploits parameters of the entire system,
in our future work, we would like to investigate convergence conditions from the local perspective, focusing on individual cluster parameters. The paper establishes fundamental concepts for this purpose, opening a new exciting avenue for future research in this field.

\section*{Appendix A: Linear Gaussian Belief Propagation}

\subsection{Message from Variable Node to Factor Node}
\begin{figure}[ht]
	\centering
	\includegraphics[width=4.0cm]{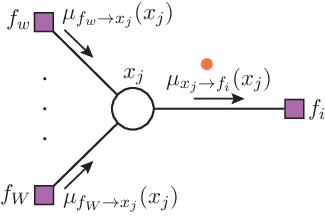}
	\caption{Message $\mu_{x_j \to f_i}(x_j)$ from variable node $x_j$ to factor node $f_i$.}
	\label{fig:fg_var_to_fac}
\end{figure}

Consider a part of a factor graph shown in~\figurename~\ref{fig:fg_var_to_fac} with a group of factor nodes $\mathcal{F}_j=\{f_i,f_w,...,f_W\}$ $\subseteq$ $\mathcal{B}$ that are neighbours of the variable node $x_j$ $\in$ $\mathcal{X}$. 
The message $\mu_{x_j \to f_i}(x_j)$ from the variable node $x_j$ to the factor node $f_i$ is equal to the product of all incoming factor node to variable node messages arriving at all the other incident edges:

\begin{equation}
    \mu_{x_j \to f_i}(x_j) =\prod_{f_a \in \mathcal{F}_j \setminus f_i} \mu_{f_a \to x_j}(x_j),
    \label{FG_v_f}
\end{equation}
where $\mathcal{F}_j \setminus f_i$ represents the set of factor nodes incident to the variable node $x_j$, excluding the factor node $f_i$. Note that each message is a function of the variable $x_j$.

Let us assume that the incoming messages $\mu_{f_w \to x_j}(x_j)$, $\dots$, $\mu_{f_W \to x_j}(x_j)$ into the variable node $x_j$ are Gaussian and represented by their mean-variance pairs $(m_{f_w \to x_j},v_{f_w \to x_j})$, $\dots$, $(m_{f_W \to x_j},v_{f_W \to x_j})$. Note that these messages carry beliefs about the variable node $x_j$ provided by its neighbouring factor nodes $\mathcal{F}_j \setminus f_i$. According to \eqref{FG_v_f}, it can be shown that the message $\mu_{x_j \to f_i}(x_j)$ from the variable node $x_j$ to the factor node $f_i$ is proportional to:  
\begin{equation}
	\mu_{x_j \to f_i}(x_j) \propto \mathcal{N}(x_j|m_{x_j \to f_i}, v_{x_j \to f_i}),		
	\label{BP_Gauss_vf} 
\end{equation}
with mean $m_{x_j \to f_i}$ and variance $v_{x_j \to f_i}$ obtained as: 
\begin{subequations}
    \begin{align}
        m_{x_j \to f_i} &= \Bigg( \sum_{f_a \in \mathcal{F}_j\setminus f_i} \cfrac{m_{f_a \to x_j}}{v_{f_a \to x_j}}\Bigg) v_{x_j \to f_i}
        \label{BP_vf_mean}\\
		\cfrac{1}{v_{x_j \to f_i}} &= \sum_{f_a \in \mathcal{F}_j\setminus f_i} \cfrac{1}{v_{f_a \to x_j}}.
	\label{BP_vf_var}
    \end{align}
	\label{BP_vf_mean_var}	
\end{subequations}

After the variable node $x_j$ receives the messages from all of the neighbouring factor nodes from the set $\mathcal{F}_j\setminus f_i$, it evaluates the message $\mu_{x_j \to f_i}(x_s)$ according to \eqref{BP_vf_mean_var} and sends it to the factor node $f_i$.

\subsection{Message from Factor Node to Variable Node}
\begin{figure}[ht]
	\centering
	\includegraphics[width=4.2cm]{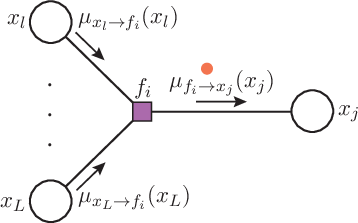}
	\caption{Message $\mu_{f_i \to x_j}(x_j)$ from factor node $f_i$ to variable node $x_j$.}
	\label{fig:fg_fac_to_var}
\end{figure}

Consider a part of a factor graph shown in~\figurename~\ref{fig:fg_fac_to_var} that consists of a group of variable nodes $\mathcal{X}_i = \{x_j, x_l,...,x_L\}$ $\subseteq$ $\mathcal X$ that are neighbours of the factor node $f_i$ $\in$ $\mathcal{B}$. 

The message $\mu_{f_i \to x_j}(x_j)$ from the factor node $f_i$ to the variable node $x_j$ is defined as a product of all incoming variable node to factor node messages arriving at other incident edges, multiplied by the local function $\mathcal{N}(z_i | \mathcal{X}_i, v_i)$ associated to the factor node $f_i$, and marginalised over all of the variables associated with the incoming messages:
\begin{equation}
    \begin{aligned}
        \mu_{f_i \to x_j}(x_j)= \int\displaylimits_{x_l}\dots\int\displaylimits_{x_L} \mathcal{N}(z_i | \mathcal{X}_i, v_i)
		\prod_{x_b \in \mathcal{X}_i\setminus x_j} \mu_{x_b \to f_i}(x_b) ~\mathrm{d}x_b, 
    \end{aligned}
	\label{FG_f_v}
\end{equation}		
where $\mathcal{X}_i\setminus x_j$ is the set of variable nodes incident to the factor node $f_i$, excluding the variable node $x_j$.

Due to linearity of functions $h_i(\mathcal{X}_i)$, closed form expressions for these messages is easy to obtain and follow a Gaussian form:
\begin{equation}
    \begin{aligned}
		\mu_{f_i \to x_j}(x_j) \propto \mathcal{N}(x_j|m_{f_i \to x_j},v_{f_i \to x_j}).
    \end{aligned}
	\label{BP_Gauss_fv}
\end{equation}
The message $\mu_{f_i \to x_j}(x_j)$ can be computed only when all other incoming messages (variable to factor node messages) are known. Let us assume that the messages into factor nodes are Gaussian, denoted by $\mu_{x_l \to f_i}(x_l) \propto \mathcal{N}(x_l|m_{x_l \to f_i}, v_{x_l \to f_i})$, $\dots$, $\mu_{x_L \to f_i}(x_L) \propto \mathcal{N}(x_L|m_{x_L \to f_i}, v_{x_L \to f_i})$. Then, using \eqref{eqn:gauss_probability} and \eqref{FG_f_v}, it can be shown that the message $\mu_{f_i \to x_j}(x_j)$ from the factor node $f_i$ to the variable node $x_j$ is represented by the Gaussian function \eqref{BP_Gauss_fv}, with mean $m_{f_i \to x_j}$ and variance $v_{f_i \to x_j}$ obtained as: 
\begin{subequations}
    \begin{align}
		m_{f_i \to x_j} &= \cfrac{1}{h_{ij}} \Bigg(z_i - \sum_{x_b \in \mathcal{X}_i \setminus x_j} 
        h_{ib} m_{x_b \to f_i} \Bigg)
        \label{BP_fv_mean}\\
        v_{f_i \to x_j} &= \cfrac{1}{h_{ij}^2} \Bigg( v_i + \sum_{x_b \in \mathcal{X}_i \setminus x_j} h_{ib}^2 v_{x_b \to f_i}  \Bigg).
		\label{BP_fv_var}
    \end{align}
	\label{BP_fv_mean_var}	
\end{subequations}

To summarise, after the factor node $f_i$ receives the messages from all of the neighbouring variable nodes from the set $\mathcal{X}_i\setminus x_j$, it evaluates the message $\mu_{f_i \to x_j}(x_j)$ according to \eqref{BP_fv_mean} and \eqref{BP_fv_var}, and sends it to the variable node $x_j$.

\subsection{Marginal Inference}
\begin{figure}[ht]
	\centering
	\includegraphics[width=3.9cm]{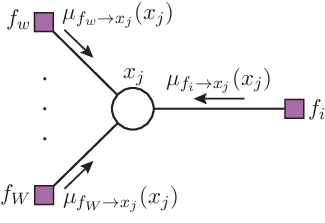}
	\caption{Marginal inference of the variable node $x_j$.}
	\label{fig:marginal}
\end{figure}

The marginal of the variable node $x_j$, illustrated in~\figurename~\ref{fig:marginal} is obtained as the product of all incoming messages into the variable node $x_j$:
\begin{equation}
    p(x_j) =\prod_{f_c \in \mathcal{F}_j} \mu_{f_c \to x_j}(x_j),
	\label{FG_marginal}
\end{equation}
where $\mathcal{F}_j$ is the set of factor nodes incident to the variable node $x_j$.		
 \noindent

According to \eqref{FG_marginal}, it can be shown that the marginal of the state variable $x_j$ is represented by: 
\begin{equation}
    p(x_j) \propto \mathcal{N}(x_j|\hat x_j,v_{x_j}),
	\label{BP_marginal_gauss}
\end{equation} 
with the mean value $\hat x_j$ and variance $v_{x_j}$:		
\begin{subequations}
    \begin{align}
        \hat x_j &= \Bigg( \sum_{f_c \in \mathcal{F}_j}
        \cfrac{m_{f_c \to x_j}}{v_{f_c \to x_j}}\Bigg) v_{x_j}
        \label{BP_marginal_mean} \\
		\cfrac{1}{v_{x_j}} &= \sum_{f_c \in \mathcal{F}_j} \cfrac{1}{v_{f_c \to x_j}}.
		\label{BP_marginal_var}        
    \end{align}
    \label{BP_marginal_mean_var}		
\end{subequations} 

Finally, the mean-value $\hat x_j$ is adopted as the estimated value of the state variable $x_j$.

\section*{Appendix B: Toy Example of Alternating Gaussian Belief Propagation Algorithm}
\begin{figure}[ht]
	\centering
	\captionsetup[subfigure]{oneside,margin={0.05cm,0cm},captionskip=5pt}
	\begin{tabular}{@{\hspace{0cm}}c@{}} 
	\subfloat[]{\label{fig:example_a}
	\includegraphics[width=3.8cm]{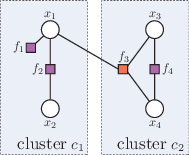}
    }
	\end{tabular} \;\;
	\captionsetup[subfigure]{oneside,margin={0.65cm,0.6cm}}
	\begin{tabular}{@{}c@{}}
	\subfloat[]{\label{fig:example_b}
    \includegraphics[width=3.8cm]{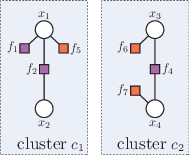}
    }
	\end{tabular}
	\caption{The complete (subfigure a) and disjoint (subfigure b) factor graph allocated to $s=2$ clusters with internal factor nodes (purple squares) and tie factor nodes (orange squares).}
	\label{fig:example}%
\end{figure}

To provide a step-by-step presentation of the AGBP algorithm, we use an illustrative example with $s=2$ clusters. The example utilises the distributed linear model given by \eqref{eqn:cluster_matrix} represented as:
\begin{equation}
    \begin{bmatrix}
        \mathbf{H}_{c_1} & \mathbf{H}_{c_1,c_2} \\
        \mathbf{H}_{c_2,c_1} & \mathbf{H}_{c_2} 
    \end{bmatrix}
    \begin{bmatrix}
        \mathbf{x}_{c_1} \\
        \mathbf{x}_{c_2} 
    \end{bmatrix} +
    \begin{bmatrix}
        \mathbf{u}_{c_1} \\
        \mathbf{u}_{c_2} 
    \end{bmatrix} =
    \begin{bmatrix}
        \mathbf{z}_{c_1} \\
        \mathbf{z}_{c_2} 
    \end{bmatrix}.
    \label{example1}
\end{equation}
Each cluster consists of two state variables:
\begin{equation}
    \mathbf{x}_{c_1} = 
    \begin{bmatrix}
        x_1 \\
        x_2 
    \end{bmatrix};\quad
        \mathbf{x}_{c_2} = 
    \begin{bmatrix}
        x_3 \\
        x_4 
    \end{bmatrix}.
    \label{example2}
\end{equation}
Furthermore, let us assume that we have the following structure of the system:
\begin{equation}
    \begin{gathered}
    \mathbf{H}_{c_1} = 
    \begin{bmatrix}
        h_{11} & 0 \\
        h_{21}  & h_{22}
    \end{bmatrix};\;
    \mathbf{H}_{c_1, c_2} = 
    \begin{bmatrix}
        0 & 0 \\
        0 & 0 
    \end{bmatrix}
    \\[3pt]
    \mathbf{H}_{c_2, c_1} = 
    \begin{bmatrix}
        h_{31} & 0 \\
        0 & 0 
    \end{bmatrix};\;
    \mathbf{H}_{c_2} = 
    \begin{bmatrix}
        h_{33} & h_{34} \\
        h_{43} & h_{44} 
    \end{bmatrix} 
    \\[3pt]
    \mathbf{z}_{c_1} = 
    \begin{bmatrix}
        z_1 \\
        z_2 
    \end{bmatrix};\;
    \mathbf{u}_{c_1} = 
    \begin{bmatrix}
        u_1 \\
        u_2 
    \end{bmatrix};\;
    \mathbf{z}_{c_2} = 
    \begin{bmatrix}
        z_3 \\
        z_4 
    \end{bmatrix};\;
        \mathbf{u}_{c_2} = 
    \begin{bmatrix}
        u_3 \\
        u_4 
    \end{bmatrix}.
        \end{gathered}
    \label{example3}
\end{equation}
Applying the AGBP algorithm to linear models \eqref{example1} requires forming the factor graph. To form a graph, we use variable nodes for representation of each state variable, resulting in the set of variable nodes $\mathcal{X}=\{x_1, x_2, x_3, x_4\}$. Similarly, each linear equation is represented by the factor node, resulting in the set of factor nodes $\mathcal{F}=\{f_1, f_2, f_3, f_4\}$, with $\mathcal{B}=\{f_2,f_3,f_4\}$ denoting branch factor nodes and $\mathcal{L}=\{f_1\}$ defining leaf factor nodes. Note that among the branch factor nodes from the set $\mathcal{B}$, only $f_3$ connects variable nodes from clusters $c_1$ and $c_2$. We label this branch factor node as the tie factor node $\mathcal{T} = \{f_3\}$. In contrast, the remaining factor nodes from the set $\mathcal{F}$ establish connections only within their respective clusters and are marked as internal factor nodes $\mathcal{I} = \{f_1,f_2,f_4\}$. \figurename~\ref{fig:example_a} depicts the resulting factor graph.

During the execution of the AGBP algorithm, each sequence $k_s = \{1, 2, \dots\}$ undergoes a series of global iterations $k_g = \{1,\dots \nu_g\}$ and local iterations $k_l = \{1,\dots \nu_l\}$. Let us assume that the algorithm starts with global iterations utilising the initial messages from the factor node to the variable node $\bm{\mu}^{(k_s,0)}_{f}$. In particular, the vectors of means $\mathbf{m}^{(k_s,0)}_{f} = [m^{(k_s,0)}_{f_i \to x_j}]$ and variances $\mathbf{v}^{(k_s,0)}_{f} = [v^{(k_s,0)}_{f_i \to x_j}]$, $f_i \in \mathcal{F}$, $x_j \in \mathcal{X}$, are known. 

We run the global iterations of the AGBP algorithm over the complete factor graph shown in \figurename~\ref{fig:example_a}. By utilising means and variances of messages $\bm{\mu}^{(k_s,k_g-1)}_{f}$, we calculate the means $\mathbf{m}^{(k_s,k_g)}_{x} = [m^{(k_s,k_g)}_{x_j \to f_i}]$ and variances $\mathbf{v}^{(k_s,k_g)}_{x} = [v^{(k_s,k_g)}_{x_j \to f_i}]$ of messages from the variable node to the factor node $\bm{\mu}^{(k_s,k_g)}_{x}$, $f_i \in \mathcal{B}$, $x_j \in \mathcal{X}$, using \eqref{BP_vf_mean_var}. For example:
\begin{subequations}
    \begin{align}
        m^{(k_s,k_g)}_{x_1 \to f_3} &= \Bigg(\cfrac{m_{f_1 \to x_1}}{v_{f_1 \to x_1}} + \cfrac{m^{(k_s,k_g-1)}_{f_2 \to x_1}}{v^{(k_s,k_g-1)}_{f_2 \to x_1}}\Bigg) v^{(k_s,k_g)}_{x_1 \to f_3}
        \\
		\cfrac{1}{v^{(k_s,k_g)}_{x_1 \to f_3}} &= \cfrac{1}{v_{f_1 \to x_1}} + \cfrac{1}{v^{(k_s,k_g-1)}_{f_2 \to x_1}}.
    \end{align}
	\label{example4}%
\end{subequations}
Note that the messages originating from the leaf factor nodes of set $\mathcal{L}$ remain constant during the entire iteration process. Next, using these messages we can compute means $\mathbf{m}^{(k_s,k_g)}_{f} = [m^{(k_s,k_g)}_{f_i \to x_j}]$ and variances $\mathbf{v}^{(k_s,k_g)}_{f} = [v^{(k_s,k_g)}_{f_i \to x_j}]$ of messages $\bm{\mu}^{(k_s,k_g)}_{f}$, $f_i \in \mathcal{B}$, $x_j \in \mathcal{X}$, using \eqref{BP_fv_mean_var}. For example:
\begin{subequations}
    \begin{align}
        m^{(k_s,k_g)}_{f_3 \to x_3} &= \cfrac{1}{h_{33}} \Bigg(z_3 - h_{31} m^{(k_s,k_g)}_{x_1 \to f_3} - h_{34} m^{(k_s,k_g)}_{x_4 \to f_3} \Bigg) \\
        v^{(k_s,k_g)}_{f_3 \to x_3} &= \cfrac{1}{h_{33}^2} \Bigg(v_3 + h_{31}^2 v^{(k_s,k_g)}_{x_1 \to f_3} + h_{34}^2 v^{(k_s,k_g)}_{x_4 \to f_3} \Bigg).
    \end{align}
    \label{example5}
\end{subequations}

Suppose that we completed the total number of global iterations $\nu_g$ for the corresponding sequence $k_s$. We can proceed with the local iterations of the AGBP algorithm on the factor graph shown in \figurename~\ref{fig:example_b}. In this graph, the tie factor node $f_3$ is collapsed into three leaf factor nodes $\{f_5,f_6,f_7\}$, each of which retains the means and variances obtained after the last global iteration and remain constant during local iterations. Following our example, the messages from these factor nodes to the corresponding variable nodes will be as follows: $\mu_{f_5 \to x_1}(x_1) = \mu^{(k_s,\nu_g)}_{f_3 \to x_1}(x_1)$, $\mu_{f_6 \to x_3}(x_3) = \mu^{(k_s,\nu_g)}_{f_3 \to x_3}(x_3)$, $\mu_{f_7 \to x_4}(x_4) = \mu^{(k_s,\nu_g)}_{f_3 \to x_4}(x_4)$. It is important to emphasise that the initial messages in the local iterations are obtained based on the results of the last global iteration. Applying the same logic as for the global iterations, we calculate the messages using equation \eqref{BP_vf_mean_var} and \eqref{BP_fv_mean_var}. The distinction lies in the fact that we only calculate messages between variable nodes from the set $\mathcal{X}$ and factor nodes from the set $\mathcal{B}\setminus \mathcal{T}$. After completing $\nu_l$ local iterations, the messages $\bm{\mu}^{(k_s,\nu_l)}_{f}$ will be used to start global iterations for a new sequence.

Our process involves repeating these sequences until the AGBP algorithm converges. Once this occurs, we can compute marginals using \eqref{BP_marginal_mean_var} according to messages from the factor node to the variable node. It is worth noting that waiting for convergence is not necessary to compute marginals. 

\section*{Appendix C: Convergence Analysis of Alternating Gaussian Belief Propagation Algorithm}
The arbitrary element of the $i$-th entry $\mathbf{c}_{f_i} = [c_{f_i \to x_j}]$, $f_i \in \mathcal{B}$, $x_j \in \mathcal{X}_{i}$, is obtained as follows: 
\begin{equation}
    c_{f_i \to x_j} = \frac{z_i}{h_{ij}} - \sum\limits_{\substack {k \in \mathcal{K} \\ h_{ik}\neq 0}} \left(\frac{h_{ik}}{h_{ij}}  \sum\limits_{\substack {r \in \mathcal{R} \\ h_{rk}\neq 0}} \frac{m_{f_r \to x_k}^*}{v_{f_r \to x_k}^*} v_{x_k \to f_i}^* \right),
    \label{eqn:bgeneric1}
\end{equation} 
where:
\begin{equation}
    v_{x_k \to f_i}^* = \left(\sum\limits_{\substack {a \in \mathcal{A} \\ h_{ak}\neq 0}} \frac{1}{v_{f_a \to x_k}^*}\right)^{-1},
    \label{eqn:bgeneric2}
\end{equation} 
with $\mathcal{K} = \{1,\dots,n \} \setminus j$, $\mathcal{R} = \{b+1,\dots, m \}$, $\mathcal{A} = \{1,\dots, m \} \setminus i$. The second part on the right-hand side of \eqref{eqn:bgeneric1} originates from the leaf factor nodes that send messages with constant mean values and variance values, representing fixed point values. 

The vector $\mathbf{m}_{f}$ can be decomposed as $\mathbf{m}_{f} = [m_1,\dots, m_q,\dots,m_d]^T$, where the mean $m_q$ corresponds to $m_{f_i \to x_j}$. Therefore, each pair of indices $(q, p)$ of the matrix $\bm {\Omega}$ corresponds to the means $(m_{f_i \to x_j}$, $m_{f_y \to x_k})$. The arbitrary element of the $i$-th block $\bm{\Omega}_{f_{i}} = [\omega_{qp}]$ can be obtained as follows:      
\begin{equation}
    \resizebox{0.89\hsize}{!}{%
    $\omega_{qp} = \begin{cases} 
      -\cfrac{h_{ik}}{h_{ij}} \cfrac{v_{x_k \to f_i}^*}{v_{f_y \to x_k}^*}, & h_{ik} \neq 0, h_{ij} \neq 0, i \neq y, j\neq k \\
      0, & \text{otherwise}.
  \end{cases}$
  }
\end{equation}

\textbf{Theorem 2 Proof}: To prove the theorem, it is sufficient to show that the sequences of global and local iterations converge to $\mathbf{m}_{f}^{*}$ as defined in \eqref{eqn:fixed_point}. Using \eqref{eqn:recursion_global} and \eqref{eqn:recursion_means_local}, we obtain the evolution of means for sequences of the global and local iterations: 
\begin{multline}
    \mathbf{m}^{(k_s, k_l)}_{f} = \mathbf{Q} {\mathbf{c}}_f + \mathbf{c}_f + \bm {\Omega} \mathbf{m}^{(k_s,\nu_g-1)}_{f} - \mathbf{Q}\mathbf{c}_f \\ -\mathbf{Q} \bm {\Omega} \mathbf{m}^{(k_s,\nu_g-1)}_{f} + \mathbf{Q}\bm {\Omega} \mathbf{m}^{(k_s,k_l-1)}_{f}.  
    \label{eqn:proof1} 
\end{multline}
Observing the fixed point $\lim_{k_s \to \infty}\mathbf{m}^{(k_s, k_l)}_{f}$, we obtained:
\begin{equation}
    \mathbf{m}_f^{*} = (\mathbf{I} - \bm{\Omega})^{-1}\mathbf{c}_f.
    \label{eqn:proof2}
\end{equation}
This concludes the proof.

\section*{Appendix D: Randomised Damping Gaussian Belief Propagation Algorithm} \label{AppendixB}
Randomised damping GBP represents an extension of the GBP algorithm, where each mean value message $m_{f_{i} \to x_{j}}$ from branch factor node $f_i \in \mathcal{B}$ to a variable node $x_j \in \mathcal{X}$ is damped independently with probability $p$, otherwise, the message is calculated as in the standard GBP algorithm. The damped message is evaluated as a linear combination of the message from the previous $m_{f_{i} \to x_{j}}^{(k-1)}$ and the current iteration step $m_{f_{i} \to x_{j}}^{(k)}$, with weights $\zeta$ and $1 - \zeta$, respectively \cite{cosovictra}:
\begin{multline}
    m_{f_{i} \to x_{j}}^{(k)}=\left(1-q_{f_i,x_j}\right) m_{f_{i} \to x_{j}}^{(k)}\\+q_{f_i,x_j} \left[(1-\zeta)  m_{f_{i} \to x_{j}}^{(k)}+ \zeta  m_{f_{i} \to x_{j}}^{(k-1)}\right], 
    \label{eqn:damping} 
\end{multline}
where $0 < \zeta < 1$, and $q_{f_i,x_j} \sim \operatorname{Ber}(p) \in\{0,1\}$ is a Bernoulli random variable independently sampled with probability $p$ for each mean $m_{f_{i} \to x_{j}}$.

\section*{Appendix E: Framework for Inference over Factor Graphs}
To ease reproducibility, we provide an open-source simulation framework written in the Julia programming language. The framework provides a powerful tool for simulation of different GBP algorithms under various scenarios, including emulation of distributed architecture of linear models, allowing straightforward use for various scenarios analysed in this paper. To emulate AGBP algorithm in the distributed architecture, we provide function \texttt{freezeFactor} which is responsible for collapsing tie factor nodes into leaf factor nodes, shown in \figurename~\ref{fig:clusters_local}. The function \texttt{defreezeFactor} returns the factor graph to its original state, shown in \figurename~\ref{fig:clusters}. For the simulation of dynamic behaviour of the GBP algorithm, we use the function \texttt{dynamicFactor}, which emulates the arrival of new observation values. Finally, we provide the function \texttt{ageingVariance}, where observations are infused with deterioration or ageing component over time. For more details, please refer to the following package \href{https://github.com/mcosovic/FactorGraph.jl}{FactorGraph.jl}, where we provide extensive documentation and source code. 

\bibliographystyle{IEEEtran}
\bibliography{cite}

\end{document}